\documentclass[reqno]{amsart}
\usepackage{xypic,graphicx}
\xyoption{all}
\usepackage{ucs}
\usepackage[utf8x]{inputenc}

\def\A{\mathcal A}
\def \Hil {\mathcal{H}}

\def \F {\mathcal{F}}

\def \D {\mathcal{D}}
\def \L{\mathcal{L}}

\def  \A{\mathcal{A}}
\def \Hil {\mathcal{H}}
\def \F {\mathcal{F}}

\def \D {\mathcal{D}}

\def \R {\mathbb{R}}

\def \C {\mathbb{C}}
\def\u {\mathfrak{u}}

\def \PH {\mathcal{PH}}
\def \RH {\mathcal{R(\Hil)}}

\def \Tr {\mathrm{Tr}}

\newcommand{\pd}[1]{\frac{\partial }{\partial #1}}

\newtheorem{corollary}{Corollary}
\newtheorem{proposition}{Proposition}
\newtheorem{definition}{Definition}
\newtheorem{lemma}{Lemma}

\newtheorem{example}{Example}
\newtheorem{remark}{Remark}

\title[Basics of Quantum Mechanics, Geometrization and some Applications \ldots]
{Basics of Quantum Mechanics, Geometrization and some  Applications to
  Quantum Information}
\author{Jes\'us Clemente-Gallardo}
\address{Instituto de Biocomputaci\'on y F\'\i sica de los Sistemas Complejos \\
  Universidad de Zaragoza \\ Corona de Arag\'on 42 \\ 50009 Zaragoza (SPAIN)}
\author{Giuseppe Marmo}
\address{Dipartamento di Scienze Fisiche \\ Universit\'a Federico II and INFN
  sezione di Napoli
  \\ Via Cintia \\ 80126 Napoli (ITALY)}

\begin{document}

\begin{abstract}
 In this paper we present a survey of the use of differential geometric
  formalisms to describe Quantum Mechanics. We analyze Schr\"odinger framework
  from this perspective and provide a description of the Weyl-Wigner
  construction. Finally, after reviewing the basics of the geometric
  formulation of quantum mechanics, we apply the methods 
  presented  to the most interesting cases of finite dimensional Hilbert
  spaces: those of two,   three and four level systems (one qubit, one qutrit
  and two qubit   systems). As a more practical application, we discuss the
  advantages that the geometric  formulation of quantum mechanics can provide
  us with in the study of situations as the   functional independence of
  entanglement witnesses. 

\end{abstract}

\maketitle
\section{Introduction}
Interference phenomena of material particles (matter waves)  as
electrons, neutrons, atoms, etc provide the most striking evidence for
the need to elaborate a new mechanics which goes beyond and
encompasses classical mechanics.

At the same time, phenomena like photoelectric and Compton effects
show a ``corpuscular'' behavior of radiation requiring, therefore, the
need for a revision of the classical description of radiation.

The associated quantitative results suggest that wave-like and
corpuscular-like attributes satisfy the following Einstein- de Broglie
relation
\begin{equation}
  \label{eq:einsteindebroglie}
  p_adx^a-Edt=\hbar (k_adx^a-\omega dt)
\end{equation}
where $(x^a, p_a)$ are the coordinates of the phase space and $t$
stands for the time. 
This relation between the Poincar\'e one form on the phase-space over
the space-time and the optical phase-space establishes a relation
between momentum ($p_a$) and  energy ($E$) of the corpuscular behavior and the
wave-number ($k_a$)  and frequency ($\omega$) of the wave
behavior. The dimensional proportionality coefficient is the Planck
constant. 

We may use this relation to predict under which experimental
conditions light will behave like a ``corpuscule'' and an electron
will behave like a ``wave''.

By means of this relation it is possible to conceive of interference
experiments like the one of the double slit. When carried on with
electrons, we find some peculiar results for which we do not have a
simple interpretation in the classical framework. The actual
experiment has been performed in such a way that at each time only one
electron is present between the source and the screen. It is found
that the electron impinges on the screen at ``given points'' and leave
a spot.

All the spots are alike, there are no ``half-spots''. After a few
hundred electrons have passed we find a picture of spots erratically
distributed on the screen, as we can see in the following picture
\begin{center}
\includegraphics[width=6cm]{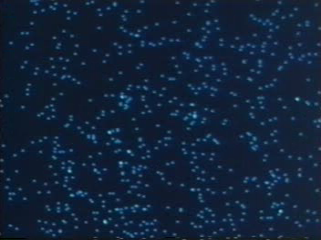}
\end{center}

 However, with several thousands electrons,
we get a very clear interference pattern.

\begin{center}
\includegraphics[width=6cm]{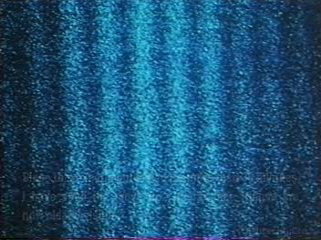}
\end{center}

Experiments of this kind have been performed (see
\cite{MerPoMini:2003,DoMiOPo:1973,DoMiOPo:1976,ToEnMAtKw:1989}). 

The same findings occur if we experiment with photons (light quanta),
again with an experimental arrangement that makes sure that only one
photon is present between source and screen at each time.

These interference aspects call for a theory with a wave-like
description, having a statistical-probabilistic character, along with
an intrinsically discrete aspect (i.e no half-spots should be possible).

We could try now to perform an experiment to watch which path the
electrons follow on its way from the source to the screen. An
experimental setup uses a conducting plate and the mirror image of the
charge to find out which region will be ``heated'' by the passage of
the electron (see \cite{Sonnetag:2006}). The setup is as follows:

\begin{center}
\includegraphics[width=10cm]{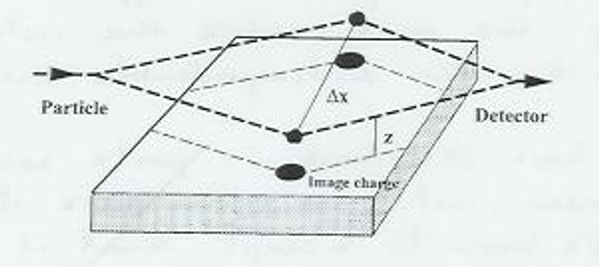}  
\end{center}

We find that for $z\approx 0$ the interference pattern disappears:

\begin{center}
\includegraphics[width=2cm]{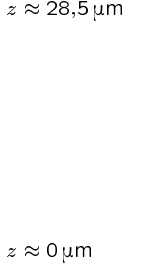}
\includegraphics[width=2cm]{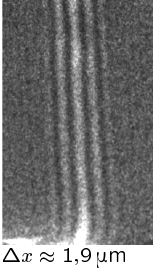} 
\includegraphics[width=2cm]{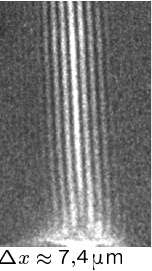}
\includegraphics[width=2cm]{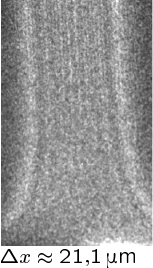} 
\end{center}

We may try to summarize the main findings of our experiments:
\begin{itemize}
\item electrons arrive at the screen in identical lumps
\item there is a wave associated with the electrons to describe the
  interference pattern
\item there is a probability for the electron to impinge on the screen
  within a preassigned region
\item under some appropriate experimental conditions we recover a
  classical-like behavior of the electrons, i.e. the interference
  pattern disappears. More specifically, the more visible the
  interference pattern is, the less distinguishable the paths are.
\end{itemize}

What we have described is quite counter-intuitive for particles, but it
is even more startling for light. We have to accept that it is not so
simple to elaborate a model capable of accounting for these results
within the classical framework. Indeed light provides the most
relevant example of the dual nature of quantum objects: its oscillatory
properties were suitable to verify the electromagnetic theory of
Maxwell, while its lumpiness (the photons) signalled the birth of
modern quantum theory. The dualism between the two pictures may appear
to be in contradiction, however it constitutes the first example of
what is known as complementarity in quantum theory, i.e. the
possibility to display both wave and corpuscular properties. 

From the historical point of view, things went differently because
inconsistencies arose already in the derivation of the spectral
distribution of energy density of a black body radiation. Max Planck
proposed  his  well known ad-hoc modification of the standard
phenomenological approach. His result, however, did not explain the
physics behind the successful result.

Another deficiency of the classical theory arose at the level of the
models available for atoms and molecules, where it was not possible to
account for the stability of atoms and molecules while explaining the
detected atomic spectra. An account of the experimental background
material can be found in most textbooks of Quantum Mechanics, for
instance in (\cite{EMS:2004}).

Various efforts of the youngest theoretical physicists of that period
gave rise to two alternative, but equivalent, formulations of quantum
mechanics, usually associated to the names of Schr\"odinger and
Heisenberg. As it is usually the case, though, the final theory grew
out of the efforts of a reasonable large community of theoretical
physicists during the first half of the XXth century.

This paper is an expanded version of some lectures delivered by one of us
(G. M.) at the ``Advanced Winter School on the Mathematical Foundation of
Quantum Control and Quantum Information'' which took place at Castro Urdiales
(Spain), February 11-15, 2008. The audience was composed of PhD students in
Mathematics and in Physics, and this explains the pedagogical nature of this
paper. The content of the lectures  has been revised and expanded in a
substantial way by both of us. 

\section{Schr\"odinger's wave mechanics}

\subsection{Introduction}
Of course, if there are waves to describe the wave-like behavior,
there ought to be a wave equation to describe the evolution of the
system in time. This equation is called the Schr\"odinger equation and
has the following structure:
\begin{equation}
  \label{eq:schrodinger}
  \frac d{dt}\psi=-\frac i\hbar H\psi. 
\end{equation}
The solutions, as functions of time, describe the time evolution of
the initial conditions of the system.  The complex valued function
$\psi$ is called {\bf the wave function} and is defined on the
configuration space of the system we are considering. From a physical
point of view it is interpreted as a probability amplitude, but this
interpretation requires, for consistency, that
\begin{equation}
  \label{eq:norm}
  \int_D \psi^*\psi d\mu=1,
\end{equation}
i.e. the probabilistic interpretation requires $\psi^* \psi d\mu$ to be
a probability density and $\psi$ to be square integrable on the domain
$D$ where our physical system is defined. We conclude thus that wave
functions representing physical systems must be elements of the
Hilbert space of square integrable functions defined on the
configuration space $D$. We shall denote this space as
$\mathcal{H}=\mathcal{L}^2(D)$. The operator $H$ is usually called the {\bf
  Hamiltonian} operator and it acts on wave functions as a linear
differential operator. The spectrum will be real, if
$H$ is self-adjoint. If it is unbounded and if its domain of
definition does not coincide with $\mathcal{H}$ we say that we have to
``face domain problems''. The requirement of linearity arises from the
superposition rule usually associated with the description of
interference phenomena.

To avoid dealing with unbounded operators, often one starts from the
``exponentiated version'' of our evolution equation, i.e. a formal
solution of the equation \ref{eq:schrodinger}, which we write as
\begin{equation}
  \label{eq:evolution}
  \psi(t)=U(t,t_0)\psi(t_0).
\end{equation}
If $H$ does not depend on time the expression of the evolution
operator $U(t,t_0)$ may be written as:
\begin{equation}
  \label{eq:ev_op}
  U(t,t_0)=e^{-\frac i \hbar H(t-t_0)}
\end{equation}

In this form the evolution operator is bounded and always
reversible. It defines a one-parameter group of unitary
transformations on the Hilbert space defined by the wave
functions. Unitarity of the evolution means that such evolution is
compatible with the probabilistic interpretation. Starting with the
evolution expressed as a one-parameter group of unitary
transformations, the operator $H$, which defines the Schr\"odinger
equation, emerges as the infinitesimal generator of the group
associated with the evolution. 

Thus we have identified the basic ingredients appearing in the
Schr\"odinger picture of Quantum Mechanics.  It is important to remark
the presence of the new fundamental constant $\hbar$ in the
description of this class of phenomena.  It implies some fundamentally
new aspects of the quantum theory with respect to the classical
theory.  For instance it is known that any measurement process
requires an exchange of energy (or information) between the object we
are measuring and the measurement apparatus. The presence of $\hbar$
implies that these exchanges can not be made arbitrarily small and
therefore idealized to be negligible. Thus, in the measurement
process, we can not conceive of a sharp separation between the
``object'' and the ``apparatus'', so that we can disregard the
apparatus altogether. Moreover, in the measuring process, there is an
inherent ambiguity in the ``cut'' between what we identify as
``object'' and what we identify as the ``apparatus''.

The problem of measurement in quantum theory is a very deep one and
goes beyond the scope of these notes. We may simple mention that
within the von Neumann formulation of Quantum Mechanics the
``measurement'' gives rise to the so called ``collapse of the wave
function''. When we measure some dynamical variable (i.e. a physical
quantity), the wave function which represents the system is projected
onto  one of the eigenstates of the operator $A$ with a probability
that can be computed. However we shall not insist on these problem
any longer and move to the general mathematical structures emerging in the
description of quantum mechanical systems.

While we shall deal mainly with finite dimensional Hilbert spaces, we
are going to consider very briefly the Weyl-Wigner formalism in
infinite dimensions and mention the tomographic formulation of
Quantum Mechanics. The reason is that this formulation is the most
suitable for the study of the quantum-classical transition.

Before closing this introduction and to better put into perspective
the Schr\"odinger and the Heisenberg pictures, we are going to make a
few general considerations on the minimal mathematical structures
required for the description of a physical system. From a minimalist
point of view, we identify three main ingredients:
\begin{itemize}
\item a space of states, which we define as $\mathcal{S}$,
\item a space of observables, which we denote as $\mathcal{O}$,
\item a real valued pairing $\mu:\mathcal{O}\times \mathcal{S}\to
  \mathbb{R}$. This pairing produces a real number out of a state and
  an observable, and it is associated with the measurement process. To
  be more precise, $\mu(A, \psi)$, for $A\in \mathcal{O}$ and $\psi\in
  \mathcal{S}$, will be a probability measure on the Borel subsets of
  $\R$. 
\end{itemize}

It is not difficult to identify these objects in the framework of
classical dynamics or of classical statistical mechanics. In Quantum
Mechanics we have two prevailing pictures:
\begin{itemize}
\item The Schr\"odinger picture. In this case $\mathcal{S}$ is
  identified with the vectors of a Hilbert space $\mathcal{H}$, while
  the set of dynamical variables (the observables) is a derived
  concept: they are identified with self-adjoint operators acting on
  $\mathcal{H}$. 
\item The Heisenberg picture. Here the situation is complementary: the
  set of dynamical variables (the observables) is the primary
  concept. Observables are assumed to be the real elements of a
  $\mathbb{C}^*$--algebra $\mathcal{A}$. The states are then a derived
  concept and are defined as a proper subset of the space of linear
  functionals on $\mathcal{A}$.
\end{itemize}

\subsection{The mathematical structure of the Schr\"odinger picture}
In this picture the relevant carrier space is a Hilbert space
$\Hil$. Very often, it is realized as a space of square-integrable
complex valued functions  defined on some spatial domain $D\subset \R^n$,
identified with the configuration space of our system.  This
configuration space turns out to be the configuration space of the
physical system that we would identify in the quantum-classical
transition when considering the so-called {\bf classical limit}
defined as $\hbar\to 0$.

States are unitary vectors, which are identified when they differ by
multiplication by a complex number of modulus one (i.e. an overall
phase). This implies an equivalence relation on $\Hil$: given $\psi_1$
and $\psi_2$ in $\Hil$ satisfying $\langle
\psi_1|\psi_1\rangle=1=\langle \psi_2|\psi_2\rangle$, we have
\begin{equation}
  \label{eq:equiv}
 \psi_2=e^{i\theta}\psi_1 \quad \theta\in [0,2\pi) \Rightarrow  \psi_1\sim \psi_2.
\end{equation}
As a result, physical states are identified with points in the complex
projective space associated to $\Hil$. We shall denote it as $\PH$ or
occasionally $\RH$ (meaning the space of rays of $\Hil$). The
corresponding equivalence classes will be denoted as
$$
  [\psi]=\{ \phi\in \Hil | \phi\sim \psi \} 
$$

On the other hand, observables are defined as self-adjoint operators
acting on $\Hil$. The pairing between states and observables is
defined in terms of the Hermitian structure of $\Hil$: with any pair
$([\psi], A)$ we associate the expectation value function $e_A$ and
defined as
\begin{equation}
  \label{eq:e_A}
  e_A([\psi])=\frac{\langle \psi, A\psi\rangle}{\langle \psi, \psi\rangle},
\end{equation}
it is clear that the right hand side depends only on the equivalence
class $[\psi]$. We are using here, as in the rest of the paper, the bra-ket
notation of Dirac \cite{Dirac:book}.

Evolution is defined on the Hilbert space $\Hil$ by means of the
Scr\"odinger equation
\begin{equation}
  \label{eq:scrod}
  i\hbar \frac{d}{dt}|\psi\rangle =H|\psi\rangle 
\end{equation}

We can elaborate a little on this equation. Consider now the equation
for the complex conjugate and transpose of the wave function:
\begin{equation}
  \label{eq:com}
   - i\hbar \frac{d}{dt}\langle  \psi| =\langle \psi| H 
\end{equation}
where the reality of $H$ has been used.

Combining both equation, it is immediate to see that, at the level of
the Hilbert space:
$$
 \frac d{dt} \langle \psi, \psi\rangle =-\frac i\hbar (\langle
  \psi, H\psi\rangle -\langle \psi H, \psi\rangle );
$$
which appears as a continuity equation.

To be specific we consider the Hilbert space realized in terms of wave
functions and the Hamiltonian operator as a differential operator
associated with a particle moving in a potential $V(\vec x)$, we find:
\begin{equation}
  \label{eq:ham_lapl}
  i\hbar \frac d{dt}\psi=-\frac{\hbar^2}{2\pi i}\Delta \psi+\hat V(\vec x)\psi,
\end{equation}
where $\Delta$ is the Laplacian operator defined on $D$ and $\hat
V(x)$ is the multiplication operator
$$
\hat V(\vec x)\psi=V(\vec x)\psi\qquad \psi\in \L^2(D)
$$

As we have mentioned, the presence of the configuration space $D$ in the
realization of the Hilbert space $\Hil$ in terms of square integrable functions
on $D$ incorporates some shadows of the classical limit. As a matter
of fact, if we introduce the polar form of the wave function as
\begin{equation}
  \label{eq:polarform}
  \psi(\vec x, t)=A(\vec x, t)e^{\frac i\hbar W(\vec x, t)},
\end{equation}
where $A$ and $W$ are real valued functions and $\psi^*\psi=A^2=\rho$,
we decompose the complex linear Schr\"odinger equation into two
nonlinear and coupled differential equations. Here the equation for
$W$ contains an additional ``quantum potential''
$$
U(\vec x, t)=-\frac {\hbar^2}{2m}\frac{\Delta A}{A},
$$
i.e. 
$$
\frac{dW}{dt}=-\frac{(\mathrm{grad} W)^2}{2m}+V(\vec x)+U(\vec x, t).
$$

The other equation takes the form of a continuity equation for the
probability distribution $\rho$:
\begin{equation}
  \label{eq:contrin}
  \frac{d\rho}{dt}+\mathrm{div} \rho \vec v=0 \qquad \vec
  v=\frac{\mathrm{grad} W}m.
\end{equation}

The vector field associated with the wave function $\psi$:
\begin{equation}
  \label{eq:bohmian}
\vec v=  \frac {d\vec x}{dt}=\frac {i\hbar}{m}\frac{\psi^* \nabla
    \psi-\psi\nabla \psi^*}{\psi^* \psi},
\end{equation}
is the extra ingredient of Bohmian mechanics (\cite{BeDurGolderZan:1995}).

Let us indulge now on the integration of the Schr\"odinger
equation. When the Hamiltonian operator does not depend explicitly on
time, i.e. it is time-translation invariant, we may use the separation
of variables and consider
$$
\psi(\vec x, t)=\phi(\vec x)e^{-\frac i\hbar Et}.
$$

We thus get the Helmholtz equation associated with the Schr\"odinger
equation:
$$
\left ( -\frac{\hbar^2}{2m}\Delta +\hat V\right ) \phi=E\phi
$$

One solves for this equation in terms of an ``eigenvalue problem''
with preassigned initial conditions (boundary conditions) and finds,
for suitable potentials,  a
set of fundamental solutions which provide also a basis for the
Hilbert space, say
$$
\phi_1, \phi_2, \cdots
$$
with the corresponding eigenvalues
$$
E_1, E_2, \cdots 
$$
A general solution for our quantum problem will thus have the form:
\begin{equation}
  \label{eq:solution}
  \psi(\vec x, t)=\sum_jc_j\phi_j(\vec x)e^{-\frac i\hbar E_jt},
\end{equation}
with $\psi(\vec x, 0)=\sum_jc_j\phi_j(\vec x)$ allowing to determine
$\{ c_j\}$ up to an overall phase. For simplicity we have considered a
situation where eigenvalues are not degenerate. The merit of the
particular basis we have considered relies on the fact that on these
vectors $\hat H$ acts as a multiplication operator, i.e. as a diagonal
matrix. Let us elaborate a little more on this aspect  in more general
terms. To this aim it is convenient to go back to the abstract Dirac
notation in terms of bras and kets. Thus starting with Schr\"odinger
equation
$$
\frac d{dt}|\psi\rangle =\frac H{i\hbar}|\psi \rangle.
$$
Let us consider now an orthonormal basis for the Hilbert space $\{
|e_i\rangle \}_{i=1, \cdots}$ and the corresponding basis for the dual
space $\{ \langle e_i|\}_{i=1, \cdots }$, with the property
$$
\langle e_j|e_i\rangle =\delta_{jk}
$$

In this basis the Hermitian structure is represented by the identity
matrix. The orthogonality also allows to write the decomposition of
the identity operator
$$
\mathbb{I}=\sum_j|e_j\rangle \langle e_j|.
$$

With the help of these bases we can write our initial Schr\"odinger
equation in the matrix form:
$$
\frac d{dt}\sum_j|e_j\rangle \langle e_j|\psi\rangle
=\sum_{jk}|e_j\rangle \langle e_j|\frac{H}{i\hbar}|e_k\rangle \langle
e_k| \psi\rangle.
$$
Now, by using the fact that $\{|e_j\rangle\}$ are a basis, we get:
$$
\frac d{dt}\psi^j=\sum_k \langle e_j|\frac{H}{i\hbar}|e_k\rangle \psi^k
$$

Denoting $h_j^k=\sum_k \langle e_j|\frac{H}{i\hbar}|e_k\rangle$ we can
write the Schr\"odinger equation as
$$
\frac d{dt}\psi^j=\sum_kh_k^j\psi^k.
$$

This equation would allow us to use all the results from matrix
algebra if the matrices would be finite dimensional. As a matter of
fact, this would be the case if for any chosen $j$ the number of
indices $k$ coming into $h_k^j$ were finite.  In geometrical terms
this would mean that we can decompose the full Hilbert space into the
direct sum of invariant finite dimensional Hilbert spaces. In this way
the infinite dimensional matrix associated with the operator would
decompose into a block diagonal form, each block being a finite
dimensional matrix.  Thus the eigenvalue problem associated with the
Helmholtz equation is nothing but a way to find such an invariant
decomposition of the Hilbert space where, in addition, each finite
dimensional block is a multiple of the identity, the multiplication
factor being the eigenvalue of the corresponding eigenspace. This
procedure gives rise to the well known spectral decomposition of the
Hamiltonian operator.

Very often, an invariant decomposition of the full Hilbert space will
be associated with the decomposition into irreducible representations
of a compact group of symmetries for the Hamiltonian operator.

In conclusion, the most convenient basis $\{ |e_i\rangle\}$ depends
on the particular problem we are dealing with, and there are no
general prescriptions.

Of course, due to the role of the observer, or apparatus, or
laboratory, a convenient basis from the point of view of
interpretation, is provided by the eigenstates of the particle's
position vector operator $\hat{\vec X}$ (whose components are the
three commuting coordinate operators $\hat x_1, \hat x_2, \hat x_3$,
defined as $\hat x_j|\vec x\rangle =x_j|\vec x \rangle$).

This continuous basis satisfies some generalized orthonormality
conditions, namely
\begin{equation}
  \label{eq:position_ortho}
  \int |\vec x\rangle d^3x\langle \vec x|=\mathbb{I} \qquad \langle
  \vec x|\vec x'\rangle =\delta(\vec x-\vec x').
\end{equation}

Decomposing $|\phi\rangle$ in the $|\vec x\rangle$ basis yields
$$
|\phi\rangle = \int |\vec x\rangle d^3x\langle \vec x|\phi\rangle.
$$

Usually we write $\langle \vec x|\phi\rangle=\phi(\vec x)$  and this
basis provides us with a clear interpretation of the wave function,
i.e. the component of the abstract vector $|\phi\rangle$ in the
basis of the position operators. Clearly the continuous  index $\vec
x$ replaces previous discrete indices $k, j$ and integration replaces
the summation. 

The position representation is quite suitable for measurements of the
particle's position. In other instances, as for instance the case of
massless particles, the momentum representation is more convenient. In
this case we use the momentum operator basis $|\vec p\rangle$
associated with the momentum operator $\hat{\vec P}$:
$$
\hat{\vec P}|\vec p\rangle =\vec p|\vec p\rangle. 
$$

Again a decomposition of the identity can be written with this basis:

\begin{equation}
  \label{eq:momentum_ortho}
  \int |\vec p\rangle d^3p\langle \vec p|=\mathbb{I} \qquad \langle
  \vec p|\vec p'\rangle =\delta(\vec p-\vec p').
\end{equation}

Each of these states can be decomposed in the position basis:
$$
|\vec p\rangle = \int |\vec x\rangle d^3x\langle \vec x|\vec
p\rangle=\frac{1}{(2\pi \hbar)^{3/2}}\int d^3xe^{\frac i\hbar \vec
  p\vec x}|\vec x\rangle 
$$

Thus the components of the momentum eigenstates, in the position
representation,  are given by plane-waves. Therefore they are not
square-integrable functions and can not be considered as elements of
$\L^2(D)$.  In this formalism, to make sense of the eigenstates
of the position or of the momentum operators, one usually deals with
wave packets. These are usually written as:
\begin{equation}
  \label{eq:packet}
  \psi(\vec x, t)=\frac 1{(2\pi \hbar)^{3/2}}\int d^3p \phi(\vec
    p)e^{\frac i\hbar(\vec p\vec x-Et)},
\end{equation}
and $E$ is written in terms of the momenta. The particular expression
of $E$ in terms of the momenta is usually called the {\bf dispersion
  relation} and the integral appearing in the momentum variables should
be considered as taken on the submanifold defined by the dispersion relation. 

The Fourier transform which appears in going from one set (of
maximally commuting) observables to a conjugate set (of maximally
commuting) variables plays a very important role because it is a
unitary transformation and therefore compatible with the probabilistic
interpretation. 

Moreover, from the mathematical point of view, we notice that the
Fourier transform represents the ``harmonic analysis'' on the Abelian
vector group. Thus it is associated with translations in space-time or
translations in the energy-momentum space \cite{varad}.

It is interesting to write now the evolution operator we mentioned
above (\ref{eq:evolution}), when we consider the position
representation. We find:
$$
\langle \vec x|\psi(t)\rangle=\int d^3x' \langle \vec x|U(t,
t_0)|\vec x'\rangle \langle x'|\psi(t_0)\rangle ,
$$
i.e.
$$
\psi(\vec x, t)=\int d^3x' G(\vec x, \vec x'; t, t_0)\psi(\vec x', t_0).
$$

Here we have introduced the standard notation for the Green function,
the ``propagator'' which makes our initial state to evolve. It simply
represents the matrix form of the evolution operator in the continuous
basis $|\vec x\rangle$ (in particular the matrix element $\langle \vec x|U(t,
t_0)|\vec x'\rangle $). It is simple to see that at the initial time
$$
G(\vec x, \vec x'; 0)=\delta(\vec x-\vec x').
$$
Therefore these matrix elements become functions on $Q\times Q$ ($Q$ being
the configuration space). By using $\int d^3x |x\rangle \langle x|$ or
$\int d^3p |p\rangle \langle p|$, we can find analogous matrix elements
written as  functions of the cotangent bundle $T^*Q$, i.e. $G(\vec
p,\vec x; t)$ or $G(\vec x ,\vec p; t)$ respectively.

The probabilistic interpretation we have considered in the
introduction is now formalized  by saying that when the state of our system
is described by the wave function $\psi(\vec x, t)$, the probability
of finding the particle in the region $D\subset \R^3$ at time $t$, is
given by
$$
\mathcal{P}(D; t)=\int_D |\psi(\vec x, t)|^2d^3x
$$

The operator 
\begin{equation}
  \label{eq:proy}
  \mathcal{P}_D=\int_D |\vec x \rangle d^3x \langle \vec x|
\end{equation}
is a projection operator and may also be described by means of the
characteristic function 
$$
\chi_D(\vec x)=
\left \{ 
\begin{array}{cr}
1 &\text{ if $\vec x\in D$} \\
0 &\text{ if $\vec x\notin D$} 
\end{array}
\right .
$$

Thus we can also write:
$$
\mathcal{P}_D=\int \chi_D(\vec x)|\vec x \rangle d^3x \langle \vec x|.
$$

Similar formulae can also be written for the momentum representation or
any other realization of the Hilbert space in terms of square
integrable functions.

Up to here, we have been using the position operator representation or
the momentum-operator representation and mentioned the possibility of
using either one. One may wonder if the general aspects
could be dealt with without making an a-priori choice of the
representation.

As a matter of fact, it is indeed possible and this is what is done in
the Weyl-Wigner formalism along with the further elaboration given by
the tomographic formalism.

The reduction to a specific representation is due to the requirement
of irreducibility of the representation of the canonical commutation
relations. In the following sections we shall briefly consider the
Weyl-Wigner formalism, the canonical commutation relations and the
emergence of these irreducible representations.

\section{Weyl systems}

\subsection{Definition and main properties}
A Weyl system is defined on a symplectic vector space $(V, \omega)$ in
the following way.  It is a map
\begin{equation}
  \label{eq:weyl}
  W: (V, \omega)\to U(\Hil), \qquad v\mapsto W(v),
\end{equation}
from the symplectic vector space to unitary operators $U(\Hil)$ on some Hilbert
space $\Hil$. We require that this map defines a  strongly continuous 
one-parameter group of transformations $\R\ni t\mapsto W(tv)$ and that it
satisfies the property
$$
W(v_1)W(v_2)=e^{\frac i\hbar \omega(v_1, v_2)}W(v_1+v_2) \quad \forall
v_1 v_2\in V
$$

Strong continuity, by means of Stone's theorem, implies the existence
of an (essentially) selfadjoint operator $R(v)$ such that
$$
W(tv)=e^{itR(v)}
$$

A well-known theorem by von Neumann establishes that for any
finite-dimensional symplectic vector space a Weyl system can always be
obtained in terms of operators acting on square-integrable functions
defined on any Lagrangian subspace $L\subset V$, with the translational
invariant Lebesgue measure $\mu_L$. We recall that a Lagrangian
subspace of a symplectic vector space is a maximal vector subspace $L$
such that $\omega(v_1, v_2)=0$ for any pair of vectors $v_1, v_2\in
L$. Maximality implies that any other vector $u\in V$ such that
$\omega(v, u)=0$ for any $v\in L$ must belong to $L$.

It is immediately seen that when vectors $v_1, v_2$ are such that
$\omega(v_1, v_2)=0$, the operators $W(v_1)$ and $W(v_2)$
commute. Therefore, a Lagrangian subspace arises as the joint spectrum
of a maximal commuting set of unitary operators.

By using vectors $|y\rangle $, eigenvectors of the operators
associated with vectors of $L$, we can write a decomposition of the
identity
$$
\mathbb{I}=\int_L |y\rangle \mu_L\langle y|,
$$
and also
$$
|\psi\rangle =\int_L |y\rangle \mu_L\langle y|\psi\rangle=\int_L
\psi(y)|y\rangle \mu_L.
$$

Within this setting, a Weyl system can be constructed in the following
way. We consider the Lagrangian subspace $L\subset V$ and its
cotangent bundle $T^*L\sim L\times L^*$, which is endowed with the canonical
symplectic form $\omega_0$.  Now we construct a symplectomorphism 
$$
(V, \omega)\longrightarrow (T^*L, \omega_0).
$$
Thus we can write the elements of $V$ as pairs $v=(x, \alpha)$ where
$x\in L$ and $\alpha\in L^*$. With this decomposition,  Weyl operators will be defined by
their action on $\L^2(L, \mu_L)$ as
$$
[W(x, 0)\psi](y)=\psi(x+y) \qquad x, y\in L \quad \psi\in \L^2(L, \mu_L)
$$
$$
[W(0, \alpha)\psi](y)=e^{i\alpha(y)}\psi(y) \qquad y\in L, \alpha\in L^* \quad
\psi\in \L^2(L, \mu_L). 
$$

We know from Stone's theorem that there exist infinitesimal generators
for these transformations. We may introduce the operators
$$
\hat P^j=R(\mathfrak{l}_j; 0)\qquad \hat Q_j=R(0; \mathfrak{l}_j),
$$
where we use $\mathfrak{l}_j$ to represent the
base vectors for $L$ and $L^*$, i.e. $\mathfrak{l}_j=(0,0, \cdots, 1,
0, \cdots, 0)$, 1 being in the $j$--th position.

With these operators, the Weyl map has the form:
\begin{equation}
  \label{eq:weyl_opera}
  W(x, \alpha)=e^{\frac i\hbar(x_j\hat P^j+\alpha^j\hat Q_j)}.
\end{equation}

Equivalently we could write instead:

$$
W_1(x, \alpha)=e^{\frac i\hbar \alpha^j\hat Q_j}e^{\frac i\hbar x_j\hat P^j},
$$
or
$$
W_2(x, \alpha)=e^{\frac i\hbar x_j\hat P^j}e^{\frac i\hbar \alpha^j\hat Q_j}.
$$

These various forms differ by multiplication by a complex number of
modulus one and give rise to different ``orderings'' (see
\cite{anmanmarsolizac:2000}). 

If we select a complex structure on $V$, defined by a tensor $J:V\to
V$ satisfying $J^2=-\mathbb{I}$, we can  define two operators

\begin{equation}
  \label{eq:creat-anhil}
a(v)=\frac 1{\sqrt{2}}(R(v)+iR(Jv)) \qquad a^+(v)==\frac 1{\sqrt{2}}(R(v)-iR(Jv)),   
\end{equation}

which will be called annihilation and creation
operators respectively. Also the natural symplectic structure
$\omega=d\alpha^j\land dx_j$ can be written in the complex
coordinates
$$
z_j=\frac 1{\sqrt{2}} (x_j+i\alpha^j) \qquad z^*_j=\frac 1{\sqrt{2}} (x_j-i\alpha^j)
$$
as
$$
\omega=idz_j^*\land dz_j.
$$
Trivially it follows that $(z_1, \cdots, z_n, \cdots )$ are
coordinates of a Lagrangian subspace, as well as  $(z^*_1, \cdots,
z^*_n, \cdots )$.

In these coordinates, we have the Weyl map given by
$$
z\mapsto D(z)=e^{za^+-z^*a}.
$$
In this form, Weyl operators are called displacement operators and are
used to construct ``coherent states''.

By means of a symplectic linear transformation, $T:V\to V$ such that
$\omega(Tv_1, Tv_2)=\omega(v_1, v_2)$, we may define an automorphism
on the set of unitary operators by setting:
$$
W(Tv)=\nu_T(W(v))=U(T)W(v)U^\dagger(T),
$$
because automorphisms of the unitary group are inner.

Thus with a one-parameter group of linear homogeneous symplectic
transformations, we define a one-parameter group of unitary
transformations. As symplectic linear transformations have an
infinitesimal generator associated with a quadratic function on $V$,
the corresponding infinitesimal generator of the unitary
transformation (Stone's theorem) shall be the operator associated with
the quadratic function on $V$. In conclusion we obtain:
\begin{lemma}
  We can associate self-adjoint operators with quadratic functions on $V$.
\end{lemma}

By using the set of complex coordinates introduced above, and the
association
$$
z\mapsto W(z)=e^{za^+-z^*a},
$$
we can construct the automorphism
$$
A\mapsto A(z)=W^\dagger(z)AW(z).
$$

By using the vacuum state $|0\rangle $ as a fiducial vector, defined
for instance as $a|0\rangle =0$, we may set a one-to-one
correspondence between Hermitian operators and functions on $V$ by
setting:
$$
f_A(z)=\langle 0|A(z)|0\rangle =\langle z|A|z\rangle.
$$
This function is known as the {\bf Berezin symbol} of the operator
$A$ and allows to define an invertible mapping between Hilbert-Schmidt
operators and square integrable functions on $V$.

We should also remark that the map
$$
z\mapsto W(z)|0\rangle=|z\rangle 
$$
is a way to immerse $\C^n$ (the phase space), as a submanifold (or a
subset) of the Hilbert space $\Hil$ (say, the Fock space). But it is
not a linear map onto the image.

If we consider the expectation value functions
$$
e_A(\psi)=\frac{\langle \psi|A|\psi\rangle}{\langle \psi|\psi\rangle},
$$
we may pull-back them to $\C^n$ by considering:
$$
e_A(z)=\frac{\langle z|A|z\rangle}{\langle z|z\rangle}.
$$

Another important point regards the algebraic structures on the
sets. As the mapping (\ref{eq:f_A}) is one-to-one, it is possible to
associate a function to the product of two operators, and thus define
a binary  operation on functions associated with operators by setting
\begin{equation}
  \label{eq:starprod}
  A\mapsto f_A(\psi), B\mapsto f_B(\psi) \quad AB\mapsto
  f_{AB}(\psi)=\langle \psi|AB|\psi\rangle=(f_A\star f_B)(\psi).
\end{equation}

When we pull-back this product to the phase space (i.e. define
$(f_A\star f_B)(z)$), we get a $star$--product very close to the usual
Moyal product (\cite{moyal}). As a matter of fact
$$
\lim_{\hbar\to 0}\frac{f_A\star f_B-f_B\star f_A}{\hbar}=\{ f_A, f_B\}, 
$$
and thus we can recover the Poisson bracket (associated to the
symplectic form $\omega$) on phase space.

Let us summarize then the results presented so far:
\begin{itemize}
\item Weyl maps are applications from a symplectic vector space $(V,
  \omega)$ to the set of unitary operators of some Hilbert space
  $U(\Hil)$ satisfying two properties:
  \begin{itemize}
  \item the map is strongly continuous,
  \item $W(v_1)W(v_2)=W(v_2)W(v_1)e^{\frac i\hbar \omega(v_1,v_2)}$
    for any $v_1, v_2\in V$.
  \end{itemize}
\item Stone's theorem allows us to define the infinitesimal generator
  of the unitary transformation corresponding to the vector $v$. We
  denote this infinitesimal generator as $iR(v)$:
$$
W(v)=e^{iR(v)}\qquad \forall v\in V.
$$
The condition above implies that
$$
[R(v_1), R(v_2)]=i\hbar \omega(v_1, v_2) \qquad \forall v_1, v_2\in V
$$

\item If we realize the symplectic vector space $V$ as the cotangent
  bundle (with its canonical symplectic form) of some Lagrangian
  submanifold $L\subset V$, i.e. $V\sim T^*L$ ; we can write the
  expression of the Weyl map as:
$$
W(x, \alpha)=e^{\frac i\hbar (x\hat P+\alpha\hat Q)}\qquad (x,
\alpha)\in T^*L;
$$
where the operators $\hat P$ and $\hat Q$ via Stone's theorem and have
an associated action on the elements of $\L^2(L, \mu_L)$ as
$$
(\hat Q\psi)(y)=y\psi(y) \quad (\hat P\psi)(y)=-i\hbar \frac{\partial
  \psi}{\partial y} \quad \psi \in \L^2(L, \mu_L).
$$
\item We can also introduce the set of complex coordinates $(z, z^*)$
  and write the expression of the Weyl map as
$$
W(z)=e^{za^+-z^*a};
$$
for $a$ and $a^+$ the creation and annihilation operators
(\ref{eq:creat-anhil}). 

\end{itemize}

\subsection{Weyl maps and Poisson tensors}
Let us elaborate a little further and
consider a similar construction, but having as initial set the Poisson
vector space $(V^*, \Lambda)$, i.e. the dual space
$V^*=\mathrm{Lin}(V, \R)$ and the Poisson tensor $\Lambda$ defined
from the symplectic form $\omega$ as:
$$
\Lambda(\alpha_{v_1}, \alpha_{v_2})=\omega(v_1, v_2),
$$
for $\alpha_{v_i}$ the element of $V^*$ defined as
$$
\alpha_{v_i}(u)=\omega(v_i, u)\qquad \forall u\in V.
$$

With these elements, we define the Weyl map as an application
$$
W:(V^*, \Lambda)\to \Hil,
$$
being strongly continuous, and satisfying the condition
$$
W(\alpha_1)W(\alpha_2)=W(\alpha_2)W(\alpha_1)e^{i\Lambda(\alpha_1, \alpha_2)}.
$$

If we use this as our starting point, we may define Weyl systems also
for degenerate Poisson structures. This turns out to be very useful
when we deal with constrained systems in the sense of Dirac and
Bergmann \cite{DuGioMarSi:1993}.

\subsection{Weyl systems and linear transformations}

Consider again the symplectic vector space $(V, \omega)$ and a linear
transformation $T:V\to V$. We can define a new tensor:
$$
\omega_T(v_1, v_2)=\omega(Tv_1, Tv_2).
$$
If $T$ is invertible, $\omega_T$ is again a symplectic form. We can
thus associate a Weyl system to the new symplectic form $\omega_T$,
in the form:
$$
\xymatrix{
(V, \omega)\ar[dd]_T\ar[rr]^W& & U(\Hil) \ar[dd]^{\nu_T} \\
\\
(V, \omega_T)\ar[rr]_{W_T} & & U(\Hil)
}
$$
As the diagram is commutative, $\nu_T(W(v))=W_T(Tv)$. 

At the infinitesimal level, we have that $R_T(v)=R(Tv)$ and therefore
$R_T(v)$ is not Hermitian with respect to the Hermitian structure on
$\Hil$ which makes $W(v)$ unitary. Hence, this is a situation where
$R_T(v)$ may have a real spectrum without being Hermitian. This
situation has been considered in the literature as the one using
pseudo-Hermitian operators (\cite{Bender:2007,Ven:2002,mastafazadeh02}). 

When $T$ is a symplectic transformation, and hence $\omega_T=\omega$,
the transformation $\nu_T$ is an inner automorphism:
$$
\nu_T(W(v))=U_T^\dagger W(v)U_T.
$$

If we consider now a one-parameter family of linear symplectic
transformations $T(\lambda)$, where $\lambda\in \R$; we know that the
infinitesimal generator of this family is a Hamiltonian vector
field. As the transformation is linear, we also know that the
associated Hamiltonian function is quadratic. But we can also consider
the infinitesimal generator of the associated unitary transformation
$U_{T(\lambda)}$. This infinitesimal generator will be a self-adjoint
operator on $\Hil$, which is thus associated to the quadratic
Hamiltonian function on $V$ which generated the family of linear
symplectic transformations. 

As an example, if we consider the transformation
$$
T=
\left (
\begin{array}{cc}
0 & -1 \\
1 & 0
\end{array}
\right )
,
$$
the associated unitary operator is given by the Fourier transform.

Let us consider now the  Heisenberg-Weyl Lie algebra $(V\oplus
\R, \omega)$ defined as
$$
[v_1, v_2]=\omega(v_1, v_2);
$$
and its holomorph, i.e. the extension of the algebra with its
derivation algebra. Let us recall that the derivation algebra of a Lie
algebra is the set of linear transformations $A:V\to V$ which satisfy:
$$
A([v_1, v_2])=[A(v_1), v_2]+[v_1, A(v_2)].
$$
In the present situation, this corresponds to the algebra of infinitesimal
canonical transformations of $V$. If we consider then this extension
of the Heisenberg-Weyl algebra, the Weyl construction shows that the
entire associated Lie group can be represented by unitary operators on
the Hilbert space on which the Weyl operators act.

\subsection{Associating operators to more general functions}
We have seen that our previous construction is able to provide a
unitary operator to represent any function of $V$ which is at most
quadratic. What can be said for more general functions?

The trick elaborated by Weyl was the following. Consider the
realization $V\sim T^*L$ which assigns a pair $(x, \alpha)$ to any
element $v\in V$. Now, we can consider the double Fourier  transform
of a function $f:V\to \R$ as
$$
\tilde f(q, p)=\int\int dxd\alpha e^{i(\alpha q+xp)} f(x, \alpha).
$$

Now define the operator $\hat f$ associated to the function $f$ as
\begin{equation}
  \label{eq:fhat}
  \hat f(\hat Q, \hat P)=\int dx d\alpha e^{\frac i\hbar(\alpha \hat
    Q+x\hat P)}f(x, \alpha).
\end{equation}

This map can be inverted, i.e. we can find a map from operators acting
on $\Hil$ onto functions defined on $V$, such that it is an
involutive map of index two. The inverse, namely the function $f_A$ associated to an
operator $A$ on the Hilbert space, can be written as:
\begin{equation}
  \label{eq:f_A}
  f_A(v)=\Tr AW(v)
\end{equation}
We might also use the symplectic Fourier transform
$$
\tilde f(q, p)=\int dx d\alpha e^{-i(\omega((x, \alpha),(q,p)))}\Tr
(\hat AW^\dagger (x, \alpha)).
$$

\begin{definition}
If we apply this to the density operator associated to a pure state $|\psi\rangle$,
$\rho_\psi=|\psi\rangle \langle \psi|$, we  obtain:
\begin{equation}
  \label{eq:Wignerfunc}
  f_{\rho_{\psi}}(q,p)=\int d\xi e^{\frac i\hbar
    p\xi}\psi(q+\xi/2)\psi^*(q-\xi/2).
\end{equation}
This is called the {\bf Wigner function} associated to the state $|\psi  \rangle$.
\end{definition}

We can conclude thus with the following statement:
\begin{lemma}
  The Wigner-Weyl map defines a bijection between Hilbert-Schmidt
  operators and square-integrable functions on phase space.
\end{lemma}

If we also denote by $W$ the map $f\mapsto \hat f=W(f)$ (i.e. applied
to functions on $V$), we can use the bijection to transport the
associative product of the $\C^*$--Weyl algebra generated by  unitary
operators on $\Hil$ onto the set of functions on $V$. This defines the
$\star$--product we saw above: 
$$
(f\star g)(x, \alpha)=W^{-1}(W(f).W(g))(x, \alpha)=\Tr \left ( W(f)W(g)W^\dagger (x,
  \alpha)\right ).
$$

This is the integral form of the Moyal product, which is usually
written in terms of bidifferential operators

\begin{equation}
  \label{eq:moyal}
  (f\star g)(q, p)=f(q, p)e^{i\frac \hbar 2\left (\stackrel{\gets}{\frac {\partial}{\partial
        q}}\stackrel{\to}{\frac {\partial}{\partial p}}-\stackrel{\gets}{\frac {\partial}{\partial
        p}}\stackrel{\to}{\frac {\partial} {\partial q}} \right )} g(q,p),
\end{equation}
or equivalently
$$
 (f\star g)(q, p)=f(q+i\frac \hbar 2\vec{\frac{\partial}{\partial p}},
 p-i\frac \hbar 2\vec{\frac{\partial}{\partial q}})g(q, p).
$$

It should be clear now that all descriptions in terms of operators
have a counterpart in terms of complex valued functions on the phase
space with the Moyal product. For instance, we can write the
eigenvalue problem for the Hamiltonian of our quantum system as:
$$
H\star \rho_E=E\rho_E,
$$
where $H$ is the function corresponding to the Hamiltonian operator
and $\rho_E$ is the function representing the eigenstate with energy
$E$. Similarly we can write evolution equations and their
exponentiation as a Taylor expansion in terms of $\star$--products.

\subsection{Tomograms}

The Wigner function associated to a state $\psi$ was introduced in 1932 as a tool
for the study of quantum corrections to classical equilibrium distributions
\cite{wigner}. However, unlike the distribution function of classical
statistical mechanics which is a probability distribution function on phase
space, the Wigner function is only a quasi-distribution, i.e. it is not
positive definite.

The tomogram description of quantum mechanics uses a formulation in which the
quantum state is described by conventional non-negative probability
distributions. Thus, within this formulation both the classical and the quantum
descriptions of a physical system use true probability distributions, and this
property makes easier to compare predictions of the two descriptions. Moreover,
it makes also easier to deal with the quantum-classical transition because the
describing functions are defined on the same carrier space.

We shall not enter in this fascinating subject, but instead refer to some
previous work to find details on this formulation in the same spirit of these
notes  \cite{manMarSiVen:2006,ManManMar:2001}

\section{The geometrical description of Quantum Mechanics}

\subsection{The Hermitian structure}
To consider Quantum Mechanics from a differential geometric point of
view, it is convenient to consider the Hilbert space $\Hil$ as a real
differential manifold. In order to do this, we have to convert the
Hermitian inner product $\langle \cdot, \cdot \rangle$ which defines
$\Hil$ into an Hermitian 
tensor. This is quite similar to what we do in the transition from
special relativity to general relativity. In that case, the Minkowski
product $\eta_{\mu\nu}x^\mu x^\nu$ is replaced by the tensor
$\eta_{\mu \nu}dx^\mu\otimes dx^\nu$ and thus space-time becomes a
manifold whose tangent space becomes endowed with a Minkowskian
product. In the present case, the Hilbert space $\Hil$ is replaced by
a Hilbertian manifold, whose tangent space is a Hilbert space with the
 inner product $\langle \cdot, \cdot \rangle$. 

To be definite, let us consider an orthonormal basis $\{ |e_j\rangle
\} _{j=1, \cdots }$ for $\Hil$ and define the coordinate functions:
$$
\langle e_j|\psi\rangle=z^j(\psi)=q^j(\psi)+ip_j(\psi),
$$
where $q^j$ and $p_j$ are real functions. With this in mind, we replace
the inner product $\langle \psi|\psi\rangle$ by  $\langle d\psi\otimes
d\psi\rangle$, where $d\psi$ is a Hilbert-space valued one-form,
i.e. a map from $T\Hil$ to $\Hil$ linear along the fibers:
$$
\sum_j |e_j\rangle \langle e_j|d\psi\rangle =\sum_jdz^j|e_j\rangle 
$$
Then the expression of the product becomes:

$$
\langle d\psi\otimes d\psi\rangle=d\bar z^j\langle e_j|e_k\rangle
dz^k=d\bar z^j\otimes dz^k\langle e_j|e_k\rangle
$$
Now using the real functions $q^j, p_j$ introduced above we get:
\begin{eqnarray*}
\langle d\psi\otimes d\psi\rangle&=& (dq^j-idp_j)\otimes
(dq^k+idp_k)\langle e_j|e_k\rangle=\\
&&\left ( dq^j\otimes dq^k+dp_j\otimes dp_k+i(dq^j\otimes
  dp_k-dp_j\otimes dq^k)  \right )\langle e_j|e_k\rangle  
\end{eqnarray*}

Because of the orthonormality of the base we find:
$$
\langle d\psi\otimes d\psi\rangle=
\left ( dq^j\otimes dq^k+dp_j\otimes dp_k+i(dq^j\otimes
  dp_k-dp_j\otimes dq^k)  \right )\delta_{jk}
$$

We can also consider the scalar product on the dual space $\Hil^*$ and
thus introduce two tensors:
\begin{equation}
  \label{eq:G}
  G=\frac{\partial}{\partial q^k}\otimes \frac{\partial}{\partial
    q_k}+\frac{\partial}{\partial p^k}\otimes \frac{\partial}{\partial
    p_k} 
\end{equation}
and
\begin{equation}
  \label{eq:lambda}
  \Lambda=\frac{\partial}{\partial p_k}\otimes \frac{\partial}{\partial
    q^k}-\frac{\partial}{\partial q^k}\otimes \frac{\partial}{\partial
    p_k} .
\end{equation}
Clearly, 
$$
\Lambda(df, dg)=\frac{\partial f}{\partial p_k}\otimes \frac{\partial g}{\partial
    q^k}-\frac{\partial f}{\partial q^k}\otimes \frac{\partial g}{\partial
    p_k} 
$$
defines a Poisson braket on the space of functions while
$$
G(df, dg)=\frac{\partial f} {\partial q^k}\otimes \frac{\partial g}{\partial
    q_k}+\frac{\partial f}{\partial p^k}\otimes \frac{\partial g}{\partial
    p_k} 
$$
defines a commutative bracket.

And we can define the Hermitian bracket as a combination of the above
products:
$$
\langle df, dg\rangle=G(df, dg)+i\Lambda(df, dg).
$$

\subsection{Other geometrical objects}

We may use the set of real coordinate functions introduced in the previous
section to write explicitly the complex structure of
$\Hil$. We know that the Hilbert space is complex, therefore if we want to
describe it as a real differentiable manifold, we need to identify the
corresponding complex structure.  It corresponds to a (1,1)--tensor
$J$, satisfying the property $J^2=-\mathbb{I}$, which written in the
coordinates above has the form
\begin{equation}
  \label{eq:J}
  J=\sum_k \left ( dq^k\otimes \frac{\partial}{\partial
      p_k}-dp_k\otimes \frac{\partial }{\partial q^k}    \right ).
\end{equation}

The other ingredient of the Hilbert space structure of $\Hil$ is its
linear structure. This also can be encoded in a tensor, a vector field
in this case, which is the dilation vector field $\Delta:\Hil\to
\Hil\times \Hil\sim T\Hil$. It is defined as $\Delta(\psi)=(\psi,
\psi)$ and hence takes the coordinate expression:
\begin{equation}
  \label{eq:delta}
  \Delta(\psi)=q^k\frac{\partial}{\partial q^k}+p_k\frac
  {\partial}{\partial p_k}
\end{equation}

The complex and the  linear structures can be thus combined as
$$
J(\Delta)=p_k\frac{\partial}{\partial q^k}-q^k\frac
  {\partial}{\partial p_k},
$$
which is the infinitesimal generator of the multiplication by a phase.

Let us go back now to the problem of the star product at the level of
the functions. Consider an operator $A$ on $\Hil$ and the evaluation functions
$$
f_A(\psi)=\langle \psi|A|\psi \rangle,
$$
and the expectation value function
$$
e_A(\psi)=\frac{\langle \psi|A|\psi \rangle}{\langle \psi|\psi\rangle}.
$$

The $\star$--product which translates the associative algebra
structure of the set of operators reads
$$
f_A\star f_B=\frac 12 G(df_A, df_B)+\frac i2 \Lambda(df_A, df_B)=f_{AB}.
$$
and
$$
e_A\star e_B=e_A e_B+\frac 12 G_P(de_A, de_B)+\frac i2 \Lambda_P(de_A, de_B)=e_{AB},
$$
where $G_P=\langle\Delta|\Delta\rangle G$ and $\Lambda_P=\langle\Delta|\Delta\rangle$ (we shall go back to these objects in the
next sections).

\begin{definition}
  A {\bf K\"ahlerian function}  is a function $f$ such that the
  Hamiltonian vector field $\Gamma_f$, 
  $\Lambda(df)=\Gamma_f$, is also a Killing vector field, i.e.
$$
L_{\Gamma_f}G=0
$$
\end{definition}

On K\"ahlerian functions, $G$ defines a Jordan algebra structure, while
$\Lambda$ defines a Lie structure. Both structures are compatible (they
define a structure called {\bf Lie-Jordan} algebra \cite{Emch,Lands:book}) and
together define the $\C^*$--algebra corresponding to bounded operators. 

The operator norm can also be obtained from the function as
$\|A\|=\mathrm{sup} e_A(\psi)$  for any positive self-adjoint operator
$A$ and any normalized vector $\psi$.

If $A$ is a general, possibly non self-adjoint operator, we have that
$\|A\|^2=\|A^*A\|$ and this is translated as
$$
\|A\|^2=\mathrm{sup}_\psi(\bar f_A\star f_A)(\psi)
$$

\begin{remark}
  The functional representation of a $\C^*$--algebra requires a
  uniform K\"ahler bundle
  \cite{CirManPizzo:1994,CirManPizzo:1991,CirManPizzo:1990I} $p:\mathcal{P}\to 
  \mathcal{B}$. $\mathcal{P}$ 
  will be the set of pure states of $\A$, while $\mathcal{B}$ is the
  spectrum of $\A$ (i.e. the set of unitary equivalent classes of
  irreducible Hilbertian representations of $\A$). $p$ is the natural
  projection, associating the equivalence class of irreducible
  representations containing the GNS representation induced by the
  pure state $\omega$:
$$
\omega_\psi(A)=e_{\pi(A)}(\psi)=\Tr \omega_\psi\pi(A) \qquad \forall A\in \A
$$
For further details on the geometrical interpretation of the GNS construction see
Chruchinski-Marmo \cite{BeppeDariusz}. 
\end{remark}

Let us consider now the relation with the notion of Gelfand transform.
The Gelfand transform of an operator $A$ is defined as
$$
e_A:\mathcal{P}\to \C \quad \mathcal{P}\ni \omega\mapsto e_A(\omega)=\omega(A)
$$
The Gelfand transform $A\mapsto e_A$ is a linear involutive preserving
injection of $\A$ into $\F(\mathcal{P})$. For $A,B\in A$,
$e_{AB}=e_A\star e_B$ and 
$$
\|A\|^2=\mathrm{sup}_{\omega\in \mathcal{P}}(\bar e_A \star e_A)(\omega).
$$
The range of the transform is the set
$$
\mathcal{K}_u(\mathcal{P})=\left \{ f\in \mathcal{K}| f, \bar f\star
  f, f\star \bar f \text{ are uniformly continuous on }
  \mathcal{P}\right \} 
$$

\subsection{Projective Hilbert spaces as K\"ahler manifolds}

We already saw above that from a physical point of view, the
probabilistic interpretation requires that the set of states of a
system is not a vector space but a complex projective one.  Our aim
now is to present the geometrical structures arising in this case.

First let us study the expression of the Hermitian tensor in $\PH$. It
is simple to obtain a Hermitian tensor field on $\Hil$ which vanishes
on vertical vector fields (generated by $\Delta$ and $J(\Delta)$) and
provides a projectable function when evaluated on projectable vector
fields:
$$
\frac{\langle d\psi\otimes d\psi\rangle}{\langle
  \psi|\psi\rangle}-\frac{\langle \psi|d\psi\rangle \otimes \langle
  d\psi|\psi\rangle}{\langle \psi|\psi\rangle^2}
$$

From a geometrical point of view, the complex projective space is
obtained as the quotient manifold obtained from the Hilbert space
$\Hil$ with respect to the foliation generated by two vector fields:
\begin{itemize}
\item the dilation vector field $\Delta=q^k\frac{\partial}{\partial
    q^k}+p_k\frac{\partial}{\partial p_k}$ , and
\item the phase-change vector field obtained as the vector field
  resulting from the action of the complex structure $J$ on $\Delta$,
  i.e. $\Gamma=J(\Delta)=q^k\frac{\partial}{\partial
    p_k}-p_k\frac{\partial}{\partial q^k}$. This vector field is the
  Hamiltonian vector field (with respect to the canonical Poisson
  structure $\Lambda$ of the function $f(q, p)=\langle
  \psi|\psi\rangle$, associated with the unity operator $\mathbb{I}$..
\end{itemize}

From this observation it is simple now to study, from a geometrical
point of view, the projection of the contravariant tensors introduced
on $\Hil$ to the complex projective space. We need to consider
contravariant tensors and functions which are invariant under the
action of $\Gamma$ and $\Delta$.

From this perspective, it is simple to see that the functions $e_A$
that we have associated with operators $A$ are trivially
projectable. On the other hand, it is simple to see (from the
coordinate expressions, for instance) that the tensors $G$ and
$\Lambda$ are homogeneous of degree -2, and hence non-projectable. The
homogeneity consideration is useful, though, to modify them by a conformal
factor and make them projectable. Thus the objects
\begin{equation}
  \label{eq:Gproj}
  G_p=\langle \Delta|\Delta\rangle G \qquad \Lambda_p=\langle
  \Delta|\Delta\rangle \Lambda,
\end{equation}
define projectable tensors.

Particularly the issue on the Poisson tensor has some deep
implications from a geometrical point of view.

We may use the two tensors above to realize the $\C^*$--algebra of
operators on $\Hil$ as a $\C^*$--algebra of functions on the complex
projective space. With the help of these functions we can formulate
the eigenvalue problem on the complex projective space considered as a
real manifold, and write the equations of motion corresponding to the
Heisenberg formulation of the evolution equations. Thanks to the
Wigner's theorem, it is possible to solve the equations of the motion
on the Hilbert space, by exponentiation for instance, and then to
project the result onto the ray space to obtain the desired
solutions. More likely it is because of this property that physicists
have not bothered dealing with Quantum Mechanics on the ray space
considered as a real manifold.

\subsection{The momentum mapping}
To end this quick description of the basics of the geometrical description of
Quantum Mechanics, we will now consider another important aspect, here worked
out for a specific example.

Given the Hilbert space $\mathcal{H}\equiv \C^3$ (or equivalently the corresponding
projective space $\PH$), we know that there is an action of the unitary group
$U(3)$ on them. With respect to the canonical symplectic structure of
$\mathcal{H}$ this action is strongly symplectic (\cite{MSSV:1985}), and therefore
admits a momentum map 
$$\mu:\mathcal{H}_0=\Hil-\{ 0\} \to \u^*(3)$$
or equivalently 
$$\tilde \mu:\PH\to 
\u^*(3)$$.

 These mappings commute with respect to the natural projection
$\pi:\mathcal{H}\to \PH$, as
$$
\xymatrix{
\Hil_0 \ar[dd]^\pi \ar[dr]^\mu & \\
& \mathfrak{u}^*(3) \\
\mathcal{P}\Hil \ar[ur]_{\tilde \mu}
}
$$

Given a point $|\psi\rangle\in \mathcal{H}$, its image by $\mu$ is an element of the dual of
the algebra of operators. From that point of view, it is easy to construct a
pairing between states and operators defined as
\begin{equation}
  \label{eq:pair}
  \mu(\psi)(A)=\langle\psi|A|\psi\rangle=\rho_\psi(A)=\Tr(A\rho_\psi)\equiv \Tr(A |\psi\rangle\langle\psi| )
\end{equation}

Thus we define the momentum mappings as
\begin{equation}
  \label{eq:mom_maps}
  \mu(\psi)\equiv \rho_\psi=|\psi\rangle\langle\psi|   \qquad  \tilde \mu(\psi)\equiv \rho_{[\psi]}=\frac{|\psi\rangle\langle\psi|}{\langle\psi|\psi\rangle},
\end{equation}
where $|\psi\rangle\langle\psi|$ is the mapping defined on $\mathcal{H}$ as
$$
|\psi\rangle\langle\psi|:\mathcal{H}\to \mathcal{H} \qquad \mathcal{H}\ni \phi \mapsto \langle\psi|\phi\rangle|\psi\rangle.
$$

Clearly the mapping $\tilde \mu$ is one-to-one. This implies that the geometric
structures that we know on $\PH$ must have an analogue on $\u^*(3)$. Indeed, it
is simple to see that the geometric structures correspond to the canonical
structures of the coadjoint orbit of $U(3)$ on $\u^*(3)$ which contains $\tilde
\mu([\psi])$.  

One important fact is the isomorphism $\u(3)\to \u^*(3)$ given by the
Killing-Cartan metric. Indeed, there is a one-to-one correspondence between the
elements of the Lie algebra and the elements of its dual and this
correspondence intertwines the adjoint and the coadjoint actions. We shall denote as
$\alpha_A$ the element of $\u^*(3)$ corresponding to the observable
$iA$ for $A\in \u(3)$.  Besides,  
as the set of physical observables in this case corresponds to the set of
linear functions on $\u^*(3)$($\u(\Hil)= \mathrm{Lin}(\u^*(\Hil), \R)$), we can
use the pullback $\mu^*:\F(\u^*(3))\to 
\F(\mathcal{H})$ to relate the set of Hermitian operators and the set of
functions defined on the Hilbert space. The geometrical structures are also
related, defining suitable morphisms:
\begin{proposition}
  \begin{itemize}
  \item[i)] The momentum map is equivariant with respect to the action of $U(\Hil)$
    on $\Hil_0$ and the coadjoint action of $U(\Hil)$ on $\mathfrak{u}^*(\Hil)$. In
    particular, this says that the Schr\"odinger equation of motion on $\Hil$ is
    $\mu$-related with the Heisenberg equation of motion on $\mathfrak{u}(\Hil)$
    (the space $\mathfrak{u}(\Hil)$ is identified with the dual by means of the
    scalar product defined by the trace). Moreover,
\item[ii)] $\mu^*(\hat A)=f_A$, $\tilde \mu^*(\hat A)=e_A$.
\item[iii)] $\mu^*(\{ \hat A, \hat B\} )=\{ f_A, f_B\} $ and $\tilde \mu^*(\{ \hat A, \hat
  B\} )=\{ e_A, e_B\} $, where 
\item[iv)] $\mu^*(R(d\hat A, d\hat B))=G(\mu^*(d\hat A), \mu^*( d\hat B))$ and for
  the other mapping  
$$\tilde \mu^*(R(d\hat A, d\hat B))=G_P(\tilde \mu^*(d\hat A),
  \tilde \mu^*( d\hat B)) +e_Ae_B, $$ 
where $R$ is the Jordan tensor defined on $\u^*(3)$ from the canonical one on
$\u(3)$ as
\begin{equation}
  \label{eq:jordantensor}
  \mathcal{R}(\xi)(d\hat A, d\hat B)=\xi([A,B]_+)=
\frac i2 \Tr \xi (AB+BA)\,,
\end{equation}
for $ \quad \xi \in \u^*(3)$, and $\hat A$ and $\hat B$ are arbitrary elements of
$\u^*(3)$. 
  \end{itemize}
\end{proposition}
\proof{
  Direct computation.
}

Instead of insisting with additional general aspects of the
geometrical formulation of Quantum Mechanics (further details may be found in
\cite{CaC-GMar:2007,CaC-GMar:2007b} , we shall consider now
some specific examples which will illustrate the general picture.

\subsection{States: Density states}
We have seen how the momentum map $\tilde \mu$ allows us to embed the complex projective
space $\mathcal P\Hil$ on the dual of the Lie algebra $\mathfrak{u}(\Hil)$. The
resulting elements represent the set of pure states of the quantum system.
But in many physical situations we have more general states, i.e. density
states which are convex combinations of pure states. They are
represented by a family $\rho=\{ \rho_1, \cdots, \rho_k\}$, each element satisfying 
$$
\rho_k^2=\rho_k,\quad \rho_k^+=\rho_k, \quad \mathrm{Tr}\rho_k=1,
$$
along with a probability vector, namely $\vec p=(p_1, p_2, \cdots , p_k)$ with
$\sum_jp_j=1$ and $p_j\geq 0 \quad \forall j$. Out of these we construct a density state
$\rho=\sum_jp_j\rho_j$. The evaluation of this state on some observable $A$ is
given by
\begin{equation}
  \label{eq:aver_mix}
\rho(A)=  \sum_jp_j \mathrm{Tr}\rho_j A=\mathrm{Tr}\rho A\,. 
\end{equation}

We shall call {\bf density states}  all convex
combinations of pure states, and we denote them by $\mathcal{D}(\Hil)$
\cite{Grabowski:2005,C-GMar:2007}.

As any of the elements in $\rho$  can be embedded into $\u^*(\mathcal{H})$, it
makes perfect 
sense to consider $\rho$ also as an element in the dual of the unitary algebra.
And we hence consider the geometric structure we defined on $\u^*(\mathcal{H})$ as the
Poisson or the Jordan brackets
\begin{eqnarray}
  \label{eq:poissonjjordan}
  \{ f_A, f_B\}(\rho) &=&\sum_kp_kf_{[A,B]_-}(\psi_k) =\sum_kp_k\{ f_A, f_B\} (\psi _k) \nonumber
  \\
(f_A, f_B)(\rho)&=&\sum_k p_k(f_A, f_B)(\psi_k)
\end{eqnarray}
where $f_A(\rho)=\sum_kp_kf_A(\psi)$.

As for the geometric structures on $\mathcal{D}(\Hil)$ we shall
 consider it as  a real manifold with boundary embedded into the real
vector space $\u^*(\mathcal{H})$. On this space the
two structures above (\ref{eq:poissonjjordan}), define a Poisson and a
``Riemannian'' structure. The Poisson structure is degenerate. However it is also
possible to define a generalized  complex structure satisfying 
\begin{equation}
  \label{eq:almost}
  J^3=-J
\end{equation}

The boundary is a stratified manifold, corresponding to the union of symplectic
orbits of $U(\mathcal{H})$  of different dimensions, passing through density matrices of
not maximal rank.  For further information see
\cite{Grabowski:2005,Grabowski:2006,ManMarSudZacc:2005}.
\section{Example I: two level quantum systems}

\subsection{The Hilbert space}
For a two levels system we shall consider an orthonormal basis on
$\C^2$, say $\{|e_1\rangle,|e_2\rangle \} $. We introduce thus a set
of coordinates
$$
\langle e_j|\psi\rangle=z^j(\psi)=q^j(\psi)+ip_j(\psi) \qquad j=1,2.
$$

In the following we shall use $z^j$ or $q^j$, $p_j$ omitting the
dependence in the state $\psi$ as it is usually done in differential
geometry.

The set of physical states is not equal to $\C^2$, since we have to
consider the equivalence relation given by the multiplication by a
complex number
i.e.
$$
\psi_1\sim \psi_2 \Leftrightarrow \psi_2=\lambda\psi_1 \qquad
\lambda\in \C_0=\C-\{ 0\}.
$$
And besides, the norm of the state must be equal to one. These two
properties can be encoded in the following diagram:
$$
\xymatrix{
\C^2\ar[rr]^\pi\ar[dr]& & S^2 \\
& S^3\ar[ur]_{\tau_H} &
}
$$
where $S^2$ and $S^3$ stand for the two and three dimensional spheres,
and the projection $\tau_H$ is the Hopf fibration.  The projection
$\pi$ is associating each vector with the one-dimensional complex
vector space to which it belongs. Thus we see how this projection
factorizes through a projection onto $S^3$ and a further projection
given by the Hopf fibration (a $U(1)$--fibration).

The Hermitian inner product on $\C^2$ can be written in the
coordinates $z_1, z_2$ as
$$
\langle \psi|\psi \rangle=\bar z_jz^k \langle e_k|e_j\rangle=\bar z_jz^j.
$$
Equivalently we can write it in real coordinates $q,p$ and obtain:
$$
\langle\psi|\psi\rangle=p_1^2+p_2^2+q_1^2+q_2^2
$$

The associated tensor field reads
$$
\langle d\psi\otimes d\psi\rangle=d\bar z_j\otimes dz^k\langle e_k |
e_j \rangle,
$$
or in real coordinates
$$
\langle d\psi\otimes d\psi\rangle=(dq_j-idp^j)\otimes
(dq^k+idp_k)\langle e_k| e_j \rangle.
$$
This expression can be decomposed into its real and its imaginary
parts and obtain
$$
dq_k\otimes dq^k+dp^k\otimes dp_k \qquad dq_j\otimes dp^j-dp^j\otimes
dq_j=dq_j\land dp^j.
$$

Hence we recognize a Riemannian structure, on the left (the real part)
and a symplectic structure (the imaginary one, on the right).

We can also obtain these tensors in contravariant form if we take as
starting point the Hilbert space $(\C^2)^*$. If we repeat the steps
above, we obtain the two contravariant tensors:
$$
G=\pd{q_k}\otimes \pd{q^k}+\pd{p^k}\otimes \pd{p_k} \qquad
\Lambda=\pd{q_k}\land \pd{p^k}.
$$

Other tensors encode the complex vector space structure of
$\Hil=\C^2$:
\begin{itemize}
\item the dilation vector field
  $\Delta=q_1\pd{q_1}+p^1\pd{p_1}+q_2\pd{q_2}+p^2\pd{p_2}$ ,
\item and the complex structure tensor $J=dp_1\otimes
  \pd{q_1}-dq^1\otimes \pd{p^1}+dp_2\otimes
  \pd{q_2}-dq^2\otimes \pd{p^2}$.
\end{itemize}

By combining both tensors, we can define the infinitesimal generator
of the multiplication by a phase:
$$
\Gamma=J(\Delta)=p_1\pd{q_1}-q^1\pd{p^1} + p_2\pd{q_2}-q^2\pd{p^2}.
$$

Thus we see how $\Delta$ is responsible for the quotienting from
$\C^2_0$ onto $S^3$, while $\Gamma$ is responsible for the Hopf
fibration $S^3\to S^2$. 

By using the Hermitian operators $\{ \sigma_0, \sigma_1, \sigma_2,
\sigma_3\}$ to construct functions $\langle \psi |A |\psi
\rangle$, we obtain the real quadratic functions
$$
q_1^2+p_1^2+q_2^2+p_2^2, \quad q_1q_2+p_1p_2, \quad q_1p_2-p_1q_2,
\quad q_1^2+p_1^2-(q_2^2+p_2^2).
$$

It is not difficult now to compute the Poisson brackets of these
quadratic functions to find that they are the Hamiltonian for the
infinitesimal generators of the $\mathfrak{u}(2)$  algebra. We may
also compute explicitly the Jordan brackets, as for instance
$$
\{q_1^2+p_1^2+q_2^2+p_2^2, q_1q_2+p_1p_2\}_+=4(q_1q_2+p_1p_2).
$$
Similar results are obtained with the other brackets. The result we
want to point out is
\begin{lemma}
  The function $q_1^2+p_1^2+q_2^2+p_2^2$ acts, with respect to the
  Jordan bracket, as the identity operator except for a normalization
  factor. 
\end{lemma}

We also find 
$$
\{ q_1q_2+p_1p_2, q_1q_2+p_1p_2\}_+=4(q_1^2+p_1^2+q_2^2+p_2^2).
$$
And analogously for the other quadratic functions. We have
\begin{lemma}
  The product of all functions (in the family above) with themselves
  produce a multiple of the quadratic isotropic function.
\end{lemma}

We also can obtain easily
$$
\{ q_1q_2+p_1p_2, q_1p_2-p_1q_2\}_+=0 =\{ q_1q_2+p_1p_2,
q_1^2+p_1^2-(q_1^2+p_1^2)\}_+ 
$$

The additional relevant property is that the Hamiltonian vector field
are Killing vectors. In terms of brackets this amounts to:
$$
\{ f, \{ g, h\} _+\} =\{ \{f, g\},h\}_+ +\{ g, \{f,h\}\}_+
$$

This condition, plus the compatibility between the associator and the
Poisson bracket implies that the two brackets combined define a
Lie-Jordan algebra.

In particular, by considering generic quadratic functions of two
complex coordinates, we find a complex valued quadratic function whose
real and imaginary parts are quadratic functions of the previous
type. All in all, the result is:
\begin{lemma}
  Complex valued quadratic functions close on a $\C^*$--algebra with respect
  to the Hermitian bracket.
\end{lemma}

Thus we have found that Hermitian operators are associated with Hamiltonian
vector fields which are also Killing. As a matter of fact, this property
characterizes functions on $\C^2$ which are associated with Hermitian
operators. 

\subsection{The projective space}
We can also consider the Hermitian tensor field related to the one on the
complex projective space. It is important to remark that while forms can not be
projected, contravariant tensor fields can. This is the reason why we
introduced the contravariant tensors $\Lambda$ and $G$. Thus by considering 
$$
G=\pd{q^k}\otimes \pd{q_k}+\pd{p_k}\otimes \pd{p_k},
$$
we can define a projectable tensor as:
\begin{equation}
  \label{eq:project_g}
  \tilde G=(q_1^2+q_2^2+p_1^2+p_2^2)\left (\pd{q^k}\otimes \pd{q_k}+\pd{p_k}\otimes
    \pd{p_k} \right)
\end{equation}

Because computations on the vector space are easier to carry on with
respect to computations on the complex projective space, it is
convenient to consider tensors on $\Hil$ which reproduce the same
results that we would have on the complex projective space. Thus
we want a tensor $\tilde G_H$ defined by
$$
\tilde G_H(\pi^*df, \pi^*dg)=\tilde G(\pi^*df, \pi^*dg),
$$
and identically vanishing on 
$$
\langle\psi|d\psi\rangle=q_jdq^j+p^jdp_j+i(q_jdp^j-p^jdq_j).
$$
The tensor is simple to construct as
\begin{equation}
  \label{eq:G_H}
  \tilde G_H=\tilde G-(\Delta\otimes \Delta+\Gamma\otimes \Gamma)-i(\Delta\otimes \Gamma-\Gamma\otimes \Delta).
\end{equation}

Analogously we can introduce
\begin{equation}
  \label{eq:lambdaH}
  \tilde \Lambda_H=(q_1^2+q_2^2+p_1^2+p_2^2)\left (\pd{q_j}\land \pd{p^j}\right )
\end{equation}

The next step is to consider the projectable quadratic functions. If we
consider the basis of the Hermitian operators given by the Pauli matrices, we
find:
$$
e_{\sigma_0}=1 \qquad e_{\sigma_1}=\frac{q_1q_2+p_1p_2}{q_1^2+q_2^2+p_1^2+p_2^2}
$$
$$
e_{\sigma_2}=\frac{q_1p_2-p_1q_2}{q_1^2+q_2^2+p_1^2+p_2^2}\qquad 
e_{\sigma_3}=\frac{q_1^2+p_1^2-q_2^2-p_2^2}{q_1^2+q_2^2+p_1^2+p_2^2}
$$

We find that only the functions associated with $\{ \sigma_1,\sigma_2,\sigma_3\}$ define
non-trivial functions on the complex projective space. Of course, their
associated vector fields generate the algebra of $SU(2)$.

We can compute now the action of the tensor $\tilde G$ on these functions, and
we obtain:
$$
\tilde G(e_{\sigma_0}, f)=0 \qquad \forall f
$$
$$
\tilde G(e_{\sigma_1}, e_{\sigma_1})=e_0-4e_{\sigma_1}^2
$$

$$
\tilde G(e_{\sigma_2}, e_{\sigma_2})=e_0-4e_{\sigma_2}^2
$$

$$
\tilde G(e_{\sigma_3}, e_{\sigma_3})=4(e_0-e_3^2)
$$

$$
\tilde G(e_{\sigma_1}, e_{\sigma_2})=-4(e_{\sigma_1}e_{\sigma_2})
$$
In an analogous way, other products can be computed. But we already see that,
due to the denominator, we do not find ``orthogonality'' of the expectation
values associated with different Pauli matrices. Instead, what we obtain 
is
\begin{lemma}
  The action of $\tilde G$ on the set of projectable functions corresponds to 
$$
\tilde G(e_A, e_B)=e_{[A, B]_+}-e_A.e_B.
$$
This implies that for $A=B$ we have
$$
\tilde G(e_A, e_B)=e_{A^2}-e_A^2,
$$
i.e. we find the variance, the quadratic deviation from the mean value. 
\end{lemma}
As a conclusion we get
\begin{corollary}
  $\tilde G$ is directly related to the indetermination relations.
\end{corollary}
\section{Example II: three level quantum systems}

\subsection{The projective space}

In the case of a three level quantum system, the Hilbert space description is
done on $\C^3$, the operators being (up to multiplication by the imaginary
unit) elements of the Lie algebra $\mathfrak{u}(3)$.

The description at the Hilbert space level  is fairly similar to the one
corresponding to the two levels 
system. Now the orthonormal basis we consider is $\{ |e_1\rangle, |e_2\rangle, |e_3\rangle\} $ and
the corresponding coordinates, complex and real will be $\{ z_1, z_2, z_3\} $
  and $\{ q_1, q_2, q_3, p_1, p_2, p_3\} $. The two restrictions:
$$
\langle\psi|\psi\rangle=1 \qquad |\psi\rangle \in \C^3
$$
and
$$
\psi_1\sim \psi_2 \Leftrightarrow \psi_2=\lambda\psi_1 \qquad \lambda\in \C,
$$
define now the corresponding sphere on the projective space $\mathcal{P}\C$.

We can read from these relations the definition of the complex and real
coordinates and the subsequent expressions for the tensors. But instead of
repeating this line of argument, we shall use a different approach and
use the momentum mapping $\mu:\mathcal{H}\to \u^*(3)$ to describe the system.

As we saw above, the relevant structures now shall be those of $\u^*(3)$, which
are defined as:
\begin{itemize}
\item the Poisson structure corresponds to the canonical Lie-Poisson structure
  of the dual of a Lie algebra, defined as:
  \begin{equation}
    \label{eq:liepoisson}
    \{ \alpha_u, \alpha_v\} =\alpha_{[u, v]} \quad \forall u,v\in \u(3), 
  \end{equation}
where we denote by $\alpha_u$ (resp. $\alpha_v$) the element of $\u^*(3)$ corresponding
to the element $u\in \u(3)$ (resp. $v\in \u(3)$).
\item the Riemannian structure is associated to the canonical Jordan structure
  of the Lie algebra $\u(3)$. Indeed, we define the tensor $R$ 
\begin{equation}
  \label{eq:jordantensor3}
  \mathcal{R}(\xi)(d\alpha_u, d\alpha_v)=\xi([u,v]_+)=
\frac i2 \Tr \xi (uv+vu)\,,
\end{equation}
\end{itemize}

Thus we see that in order to compute the expression of these tensors in some
basis, we have to compute the structure constants of the $\u(3)$ algebra. Thus
if we take as basis the one defined by the Gell-Mann matrices

 $$
\lambda_1=
\left (
\begin{array}{ccc}
0 & 1 & 0 \\
1 & 0 & 0 \\
0 & 0 & 0
\end{array}
\right )
\quad 
\lambda_2=
\left (
\begin{array}{ccc}
0 & -i & 0 \\
i & 0 & 0 \\
0 & 0 & 0
\end{array}
\right )
\quad 
\lambda_3=
\left (
\begin{array}{ccc}
1 & 0 & 0 \\
0 & -1 & 0 \\
0 & 0 & 0
\end{array}
\right )
$$
$$ 
\lambda_4=
\left (
\begin{array}{ccc}
0 & 0 & 1 \\
0 & 0 & 0 \\
1 & 0 & 0
\end{array}
\right )
\quad
\lambda_5=
\left (
\begin{array}{ccc}
0 & 0 & -i \\
0 & 0 & 0 \\
i & 0 & 0
\end{array}
\right )
\quad 
\lambda_6=
\left (
\begin{array}{ccc}
0 & 0 & 0 \\
0 & 0 & 1 \\
0 & 1 & 0
\end{array}
\right )
$$
$$ 
\lambda_7=
\left (
\begin{array}{ccc}
0 & 0 & 0 \\
0 & 0 & -i \\
0 & i & 0
\end{array}
\right )
\quad 
\lambda_8=\frac 1{\sqrt{3}}
\left (
\begin{array}{ccc}
1 & 0 & 0 \\
0 & 1 & 0 \\
0 & 0 & -2
\end{array}
\right )
\quad 
\lambda_0=\sqrt{\frac 2 3}
\left (
\begin{array}{ccc}
1 & 0 & 0 \\
0 & 1 & 0 \\
0 & 0 & 1
\end{array}
\right )
$$

They satisfy the scalar product relations
$$
\Tr \lambda_\mu \lambda_\nu=2\delta_{\mu \nu}
$$

Their commutation and anti-commutation relations are written in terms of the
antisymmetric structure constants and symmetric d--symbols $d_{\mu \nu \rho}$. We find
$$
[\lambda_\mu, \lambda_\nu]=2iC_{\mu \nu \rho}\lambda_\rho \quad
[\lambda_\mu, \lambda_\rho]_+=2\sqrt{\frac 2 3} \lambda_0\delta_{\mu \nu}+2d_{\mu \nu
  \rho}\lambda_\rho. 
$$

The numerical values turn out to be
$$
C_{123}=1, \qquad C_{458}=C_{678}=\frac{\sqrt{3}}2 
C_{147}=-C_{156}=C_{246}=C_{257}=C_{345}=-C_{367}=\frac 12
$$
The values of these symbols show the different embeddings of $SU(2)$ into
$SU(3)\subset U(3)$.  For the other coefficients we have

\begin{eqnarray*}
&d_{jj0}=-d_{0jj}=-d_{j0j}=\sqrt{\frac 23} \quad j=1, \cdots , 8\\
&-d_{888}=d_{jj8}=d_{j8j}=\frac 1{\sqrt{3}}\quad j=1, 2, 3 \\
&d_{8jj}=d_{jj8}=d_{j8j}=-\frac 1{2\sqrt{3}}  \quad j=4,5,6,7\\
&d_{3jj}=d_{jj3}=d_{j3j}=\frac 1{2}  \quad j=4,5 \qquad
d_{3jj}=d_{jj3}=d_{j3j}=-\frac 1{2}  \quad j=6,7\\
&d_{146}=d_{157}=d_{164}=d_{175}=-d_{247}=d_{256}=d_{265}=-d_{274}=\frac 12 \\
&d_{416}=-d_{427}=d_{461}=-d_{472}=d_{517}=d_{526}=d_{562}=d_{571}=\frac 12 \\
&d_{614}=d_{625}=d_{641}=d_{652}=d_{715}=-d_{724}=d_{751}=-d_{742}=\frac 12
\end{eqnarray*}

Thus we have completely characterized  the Riemannian and the Poisson tensors
along with their associated brackets on the density states of the
three-levels system. 
\section{Example III: a bipartite composite system of two spin $\frac
  12$ particles}
Our last example will be devoted to the study of a system composed of two
spin-$1/2$ particles.  In this case the Hilbert space is the 4 dimensional
complex space $\C^4=\C^2\otimes \C^2$ and the set of Hermitian operators corresponds to
the elements of the Lie algebra $\u(4)$ after multiplication by the
imaginary unit. 

The importance of the example arises from the fact that this is the simplest
situation where entanglement of two systems can be studied.

We will use then the approach used in the previous section, and describe the
geometrical structure of the composite system by using the momentum map and
thus the image of the Hilbert space $\C^4$ on $\u^*(4)$.

\subsection{The choice of coordinates on $\u^*(4)$}
The
 most suitable choices corresponds to the tensor product of the basis
 of the two $\u(2)$ factors corresponding to each subsystem: 
\begin{equation}
 \mathcal{B}=\left \{ \sigma_k\otimes \tau_j \right \}_{k,j=0,1,2,3},
\end{equation} 
where $\sigma_0$ stands for the identity matrix in dimension 2 and
$\{\sigma_i\}_{i=1,2,3}$ and $\{\tau_i\}_{i=1,2,3}$ represent two
copies of the Pauli matrices. Thus we represent the elements in
$\u(4)$ by a set of 16 real coordinates $\{\lambda_0,
m_k,n_j,r_{kj}\}$ for $k,j=1,2,3$ defined as: 
\begin{equation}
\u(4)\ni A=\lambda_0\sigma_0\otimes \tau_0+\sum_{k=1}^3m_k\sigma_k\otimes \tau_0+\sum_{k=1}^3n_k\sigma_0\otimes
\tau_k+\sum_{k,j=1}^3(m_kn_j+r_{kj})\sigma_k\otimes \tau_j 
\end{equation} 

Now all tensor fields shall be written in terms of these coordinates
in the next sections. This basis has an immediate extension to
$\u^*(4)$ by the duality relation defined by the invertibility of the
Killing-Cartan form. The main advantage of this choice of coordinates
is that separable elements in the space of density operators
correspond to the points defined by setting 
\begin{equation}
 r_{ij}=0 \qquad \forall i,j
\end{equation} 

\subsection{The Riemann and the Poisson structures}

The computation of the Poisson and the Riemann tensor requires again to compute
the structure constants of the Lie and the Jordan algebras in this basis. Let
us present the results by means of the corresponding Hamiltonian and Riemannian
vector fields associated to the basis elements:

\subsubsection{Hamiltonian vector fields}
Besides the
Hamiltonian vector field associated to $\lambda_0$, which is a Casimir
of the structure, these are the Hamiltonian vector fields associated
to the elements of the basis. We list just a few of them, since the others can be obtained from these ones by a simple permutation of indices
 
\small
\begin{eqnarray*}
 \{m_1, \cdot\}&=&
r_{33}\pd{r_{23}}
-r_{23}\pd{r_{33}}+r_{32}\pd{r_{22}}-r_{22}\pd{r_{32}}+ \\
&&r_{21}\pd{r_{31}}
-r_{31}\pd{r_{21}}-m_{2}\pd{m_{3}}   
 +m_{3}\pd{m_{2}}\\ 
\{n_1, \cdot\}&=&
-r_{32}\pd{r_{33}}+r_{33}\pd{r_{32}}-r_{22}\pd{r_{23}}+r_{23}\pd{r_{22}}-\\
&&r_{12}\pd{r_{13}}+r_{13}\pd{r_{12}}-n_{2}\pd{n_{3}}
+n_{3}\pd{n_{2}}\\
\{r_{11}, \cdot \}&=&
 r_{12}\pd{n_3}- \left(\left(-1+m_1^2\right) n_2+2 m_1
  r_{12}\right)\pd{r_{13}}-\\
&&r_{13}\pd{n_2} +\left(\left(-1+m_1^2\right) n_3+2
  m_1 r_{13}\right)\pd{r_{12}} + \\ 
&& r_{21}\pd{m_3}-\left(m_2 \left(-1+n_1^2\right)+2 n_1
  r_{21}\right)\pd{r_{31}} -\\
&&r_{31}\pd{m_2} +\left(m_3 \left(-1+n_1^2\right)+2
  n_1 r_{31}\right)\pd{r_{21}} + \\ 
&& \left(m_2 r_{13}+m_1 \left(m_2 n_3+r_{23}\right)+n_2 \left(m_3
    n_1+r_{31}\right)+n_1 r_{32}\right)\pd{r_{22}}- \\ 
&& \left(m_3 r_{12}+n_3 \left(m_2 n_1+r_{21}\right)+n_1 r_{23}+m_1
  \left(m_3 n_2+r_{32}\right)\right)\pd{r_{33}}+ \\ 
&& \left(-m_2 r_{12}-m_1 \left(m_2 n_2+r_{22}\right)+n_3 \left(m_3
    n_1+r_{31}\right)+n_1 r_{33}\right)\pd{r_{23}} \\ 
&&+ \left(m_3 r_{13}-n_2 \left(m_2 n_1+r_{21}\right)-n_1 r_{22}+m_1
  \left(m_3 n_3+r_{33}\right)\right)\pd{r_{32}} \\
\{r_{12}, \cdot\}&=& 
r_{11}\pd{n_3} - \left(\left(-1+m_1^2\right) n_1+2 m_1
  r_{11}\right)\pd{r_{13}}-\\
&& r_{13}\pd{n_1}+ \left(\left(-1+m_1^2\right) n_3+2
  m_1 r_{13}\right)\pd{r_{11}} \\ 
&&- r_{22}\pd{m_3}+ \left(m_2 \left(-1+n_2^2\right)+2 n_2 r_{22}\right)\pd{r_{32}}+ \\
&& \left(-m_3 r_{11}+n_3 \left(m_2 n_2+r_{22}\right)+n_2 r_{23}-m_1
  \left(m_3 n_1+r_{31}\right)\right)\pd{r_{33}}+r_{32}\pd{m_2} + \\ 
&& \left(m_2 r_{13}+m_1 \left(m_2 n_3+r_{23}\right)-n_2 \left(m_3
    n_1+r_{31}\right)-n_1 r_{32}\right)\pd{r_{21}}- \\ 
&& \left(m_3 \left(-1+n_2^2\right)+2 n_2 r_{32}\right)\pd{r_{22}}- \\
&& \left(m_2 r_{11}+m_1 \left(m_2 n_1+r_{21}\right)+n_3 \left(m_3
    n_2+r_{32}\right)+n_2 r_{33}\right)\pd{r_{23}}+ \\ 
&& \left(m_3 r_{13}+n_2 \left(m_2 n_1+r_{21}\right)+n_1 r_{22}+m_1
  \left(m_3 n_3+r_{33}\right)\right)\pd{r_{31}} 
\end{eqnarray*}
\begin{eqnarray*}
\{r_{21}, \cdot\}&=& 
r_{11}\pd{m_3} -\left(m_1 \left(-1+n_1^2\right)+2 n_1
  r_{11}\right)\pd{r_{31}} -\\
&& r_{22}\pd{n_3}+ \left(\left(-1+m_2^2\right) n_2+2
  m_2 r_{22}\right)\pd{r_{23}}+ \\ 
&&r_{23}\pd{n_2} - \left(\left(-1+m_2^2\right) n_3+2 m_2
  r_{23}\right)\pd{r_{22}}-\\
&&r_{31}\pd{m_1} + \left(m_3 \left(-1+n_1^2\right)+2
  n_1 r_{31}\right)\pd{r_{11}}+ \\ 
&& \left(-m_2 r_{13}-m_1 \left(m_2 n_3+r_{23}\right)+n_2 \left(m_3
    n_1+r_{31}\right)+n_1 r_{32}\right)\pd{r_{12}}+ \\ 
&& \left(-n_3 \left(m_1 n_1+r_{11}\right)-n_1 r_{13}+m_3 r_{22}+m_2
  \left(m_3 n_2+r_{32}\right)\right)\pd{r_{33}}+ \\ 
&& \left(m_2 r_{12}+m_1 \left(m_2 n_2+r_{22}\right)+n_3 \left(m_3
    n_1+r_{31}\right)+n_1 r_{33}\right)\pd{r_{13}}- \\ 
&& \left(n_2 \left(m_1 n_1+r_{11}\right)+n_1 r_{12}+m_3 r_{23}+m_2
  \left(m_3 n_3+r_{33}\right)\right)\pd{r_{32}} \\ 
\end{eqnarray*} 
 
\subsubsection{Riemannian vector fields}
Also by using the Riemannian structure we can associate vector fields with
coordinate functions. Again, we just list a few of them: 
\begin{eqnarray*}
J_{\lambda_0}&=&
n_ 1 \frac{\partial}{\partial n_ 1} + n_ 2 \frac{\partial}{\partial n_ 2} + n_
3 \frac{\partial}{\partial n_ 3} + 
m_ 1 \frac{\partial}{\partial m_ 1} + m_ 2 \frac{\partial}{\partial m_ 2} + m_
3 \frac{\partial}{\partial m_ 3} + 
y_ 0 \frac{\partial}{\partial y_ 0} +\\
&& r_ {11} \frac{\partial}{\partial r_ {11}} + r_ {12} \frac{\partial}{\partial r_ {12}} +
 r_ {13} \frac{\partial}{\partial r_ {13}} +
 r_ {21} \frac{\partial}{\partial r_ {21}} + r_ {22} \frac{\partial}{\partial r_ {22}} +
 r_ {23} \frac{\partial}{\partial r_ {23}} +\\
&& r_ {31} \frac{\partial}{\partial r_ {31}} + r_ {32} \frac{\partial}{\partial r_ {32}} +
 r_ {33} \frac{\partial}{\partial r_ {33}}  
 - n_ 1m_1 \frac{\partial}{\partial r_ {11}} -
    n_ 2m_1 \frac{\partial}{\partial r_ {12}} - n_3m_1
    \frac{\partial}{\partial r_ {13}}  +\\ 
 &&- n_ 1m_2
 \frac{\partial}{\partial r_ {21}} -
    n_ 2 m_2\frac{\partial}{\partial r_ {22}} - n_3m_2
    \frac{\partial}{\partial r_ {23}}  + 
 - n_1m _3\frac{\partial}{\partial r_ {31}} -
    n_ 2 m_3\frac{\partial}{\partial r_ {32}} - n_3m_3
    \frac{\partial}{\partial r_ {33}}   
\\
J_{m_1}&=&
y_ 0 \frac{\partial}{\partial m_
    1} + \left (m_ 1 n_ 1 + r_ {11} \right) \frac{\partial}{\partial n_
    1} + \left (m_ 1 n_ 2 + r_ {12} \right) \frac{\partial}{\partial n_
    2} + \left (m_ 1 n_ 3 + r_ {13} \right) \frac{\partial}{\partial n_
    3} -\\
&&
 \left (m_ 1 r_ {11} + 
     n_ 1 \left (m_ 1^2 + y_ 0 - 1 \right) \right) \frac{\partial}{\partial r_ {11}} - \left
     (m_ 1 r_ {12} + 
     n_ 2 \left (m_ 1^2 + y_ 0 - 1 \right) \right) \frac{\partial}{\partial r_ {12}} -\\
&& \left (m_ 1 r_ {13} +
     n_ 3 \left (m_ 1^2 + y_ 0 - 1 \right) \right) \frac{\partial}{\partial r_ {13}} -\\
&& m_ 2 \left (m_ 1 n_ 1 + r_ {11} \right) \frac{\partial}{\partial r_ {21}} -
 m_ 2 \left (m_ 1 n_ 2 + r_ {12} \right) \frac{\partial}{\partial r_ {22}} -\\
&& m_ 2 \left (m_ 1 n_ 3 + r_ {13} \right) \frac{\partial}{\partial r_ {23}} -
 m_ 3 \left (m_ 1 n_ 1 + r_ {11} \right) \frac{\partial}{\partial r_ {31}} -\\
&& m_ 3 \left (m_ 1 n_ 2 + r_ {12} \right) \frac{\partial}{\partial r_ {32}} -
 m_ 3 \left (m_ 1 n_ 3 + r_ {13} \right) \frac{\partial}{\partial r_ {33}} +\\
&& m_ 1 \frac{\partial}{\partial y_ 0} 
\end{eqnarray*}
\begin{eqnarray*}
 J_{n_1}&=&
\left (m_ 1 n_ 1 + r_ {11} \right) \frac{\partial}{\partial m_
    1} + \left (m_ 2 n_ 1 + r_ {21} \right) \frac{\partial}{\partial m_
    2} + \\
&& \left (m_ 3 n_ 1 + r_ {31} \right) \frac{\partial}{\partial m_ 3} +
 y_ 0 \frac{\partial}{\partial n_
    1} - 
\left (n_ 1 r_ {11} + 
     m_ 1 \left (n_ 1^2 + y_ 0 - 1 \right) \right) \frac{\partial}{\partial r_ {11}} - \\
&& n_ 2 \left (m_ 1 n_ 1 + r_ {11} \right) \frac{\partial}{\partial r_ {12}} -
 n_ 3 \left (m_ 1 n_ 1 + 
     r_ {11} \right) \frac{\partial}{\partial r_ {13}} - \\
&&\left (n_ 1 r_ {21} +
     m_ 2 \left (n_ 1^2 + y_ 0 - 1 \right) \right) \frac{\partial}{\partial r_ {21}} -
 n_ 2 \left (m_ 2 n_ 1 + r_ {21} \right) \frac{\partial}{\partial r_ {22}} -
 n_ 3 \left (m_ 2 n_ 1 + 
     r_ {21} \right) \frac{\partial}{\partial r_ {23}} -\\
&& \left (n_ 1 r_ {31} +
     m_ 3 \left (n_ 1^2 + y_ 0 - 1 \right) \right) \frac{\partial}{\partial r_ {31}} -
 n_ 2 \left (m_ 3 n_ 1 + r_ {31} \right) \frac{\partial}{\partial r_ {32}} -
 n_ 3 \left (m_ 3 n_ 1 + r_ {31} \right) \frac{\partial}{\partial r_ {33}} +
 n_ 1 \frac{\partial}{\partial y_ 0} \\
J_{r_{11}}&=&
\left (m_ 1 r_ {11} + 
     n_ 1 \left (m_ 1^2 + y_ 0 - 1 \right ) \right ) \frac{\partial}{\partial m_ 1} +
  m_ 1 \left (m_ 2 n_ 1 + r_ {21} \right ) \frac{\partial}{\partial m_ 2} +\\
&&
 m_ 1 \left (m_ 3 n_ 1 + r_ {31} \right ) \frac{\partial}{\partial m_
    3} + \left (n_ 1 r_ {11} + 
     m_ 1 \left (n_ 1^2 + y_ 0 - 1 \right ) \right ) \frac{\partial}{\partial n_ 1} +\\
 && n_ 1 \left (m_ 1 n_ 2 + r_ {12} \right ) \frac{\partial}{\partial n_ 2} +
 n_ 1 \left (m_ 1 n_ 3 + r_ {13} \right ) \frac{\partial}{\partial n_
    3} -\\
&&\left (\left (2 n_ 1^2 + y_ 0 - 2 \right ) m_ 1^2 +
     2 n_ 1 r_ {11} m_ 1 + n_ 1^2 \left (y_ 0 - 2 \right ) +
     y_ 0 \right ) \frac{\partial}{\partial r_ {11}} -\\
&&\left (m_ 1 n_ 2 r_ {11} +
     n_ 1 \left (m_ 1 r_ {12} + 
        n_ 2 \left (2 m_ 1^2 + y_ 0 - 2 \right ) \right ) \right) \frac{\partial}{\partial r_
      {12}} - \\
&&
\left (m_ 1 n_ 3 r_ {11} +
     n_ 1 \left (m_ 1 r_ {13} + 
        n_ 3 \left (2 m_ 1^2 + y_ 0 - 2 \right ) \right ) \right)
    \frac{\partial}{\partial r_ {13}} -\\
&&
 \left (m_ 2 n_ 1 r_ {11} +
     m_ 1 \left (n_ 1 r_ {21} + 
        m_ 2 \left (2 n_ 1^2 + y_ 0 - 2 \right ) \right ) \right)
    \frac{\partial}{\partial r_ {21}} +\\
&&
\left (m_ 3 n_ 3 -
     m_ 2 n_ 1 r_ {12} - 
     m_ 1 n_ 2 \left (2 m_ 2 n_ 1 + r_ {21} \right ) +
     r_ {33} \right ) \frac{\partial}{\partial r_ {22}} -\\
&&
 \left (m_ 3 n_ 2 +
     m_ 2 n_ 1 r_ {13} + 
     m_ 1 n_ 3 \left (2 m_ 2 n_ 1 + r_ {21} \right ) +
     r_ {32} \right ) \frac{\partial}{\partial r_ {23}} -\\
&&
\left (m_ 3 n_ 1 r_ {11} +
     m_ 1 \left (n_ 1 r_ {31} + 
        m_ 3 \left (2 n_ 1^2 + y_ 0 - 2 \right ) \right ) \right)
    \frac{\partial}{\partial r_ {31}} -\\
&&
 \left (m_ 2 n_ 3 + 
     m_ 3 n_ 1 r_ {12} + r_ {23} + 
     m_ 1 n_ 2 \left (2 m_ 3 n_ 1 + r_ {31} \right ) \right)
   \frac{\partial}{\partial r_ {32}} +\\
&&
 \left (m_ 2 n_ 2 -
     m_ 3 n_ 1 r_ {13} + r_ {22} - 
     m_ 1 n_ 3 \left (2 m_ 3 n_ 1 + r_ {31} \right ) \right) \frac{\partial}{\partial r_ {33}} +
\left (m_ 1 n_ 1 -
     r_ {11} \right ) \frac{\partial}{\partial y_ 0} 
\end{eqnarray*}
\begin{eqnarray*}
J_{r_{12}}&=&
\left (m_ 1 r_ {12} + 
     n_ 2 \left (m_ 1^2 + y_ 0 - 1 \right ) \right ) \frac{\partial}{\partial m_ 1} +
  m_ 1 \left (m_ 2 n_ 2 + r_ {22} \right ) \frac{\partial}{\partial m_ 2} +\\
&&
 m_ 1 \left (m_ 3 n_ 2 + r_ {32} \right ) \frac{\partial}{\partial m_ 3} +
 n_ 2 \left (m_ 1 n_ 1 + r_ {11} \right ) \frac{\partial}{\partial n_
    1} +\\
&&
 \left (n_ 2 r_ {12} + 
     m_ 1 \left (n_ 2^2 + y_ 0 - 1 \right ) \right ) \frac{\partial}{\partial n_ 2} +
  n_ 2 \left (m_ 1 n_ 3 + r_ {13} \right ) \frac{\partial}{\partial n_
    3} -\\
&&
 \left (m_ 1 n_ 2 r_ {11} + 
     n_ 1 \left (m_ 1 r_ {12} + 
        n_ 2 \left (2 m_ 1^2 + y_ 0 - 2 \right ) \right ) \right)
    \frac{\partial}{\partial r_ {11}} -\\
&&
\left (\left (2 n_ 2^2 + y_ 0 -
        2 \right ) m_ 1^2 + 2 n_ 2 r_ {12} m_ 1 +
     n_ 2^2 \left (y_ 0 - 2 \right ) +
     y_ 0 \right ) \frac{\partial}{\partial r_ {12}} -\\
&&
\left (m_ 1 n_ 3 r_ {12} +
     n_ 2 \left (m_ 1 r_ {13} + 
        n_ 3 \left (2 m_ 1^2 + y_ 0 - 2 \right ) \right ) \right)
    \frac{\partial}{\partial r_ {13}} -\\
&&
 \left (m_ 3 n_ 3 +
     m_ 2 n_ 2 r_ {11} + 
     m_ 1 n_ 1 \left (2 m_ 2 n_ 2 + r_ {22} \right ) +
     r_ {33} \right ) \frac{\partial}{\partial r_ {21}} -\\
&&
\left (m_ 2 n_ 2 r_ {12} +
     m_ 1 \left (n_ 2 r_ {22} + 
        m_ 2 \left (2 n_ 2^2 + y_ 0 - 2 \right ) \right ) \right) \frac{\partial}{\partial r_
      {22}} + \\
&&
\left (m_ 3 n_ 1 -
     m_ 2 n_ 2 r_ {13} - 
     m_ 1 n_ 3 \left (2 m_ 2 n_ 2 + r_ {22} \right ) +
     r_ {31} \right ) \frac{\partial}{\partial r_ {23}} +\\
&&
 \left (m_ 2 n_ 3 -
     m_ 3 n_ 2 r_ {11} + r_ {23} - 
     m_ 1 n_ 1 \left (2 m_ 3 n_ 2 + r_ {32} \right ) \right)
   \frac{\partial}{\partial r_ {31}} -\\
&&
\left (m_ 3 n_ 2 r_ {12} +
     m_ 1 \left (n_ 2 r_ {32} + 
        m_ 3 \left (2 n_ 2^2 + y_ 0 - 2 \right ) \right ) \right)
    \frac{\partial}{\partial r_ {32}} -\\
&&
 \left (m_ 2 n_ 1 +
     m_ 3 n_ 2 r_ {13} + r_ {21} + 
     m_ 1 n_ 3 \left (2 m_ 3 n_ 2 + r_ {32} \right ) \right) \frac{\partial}{\partial r_
     {33}} + \left (m_ 1 n_ 2 - 
     r_ {12} \right ) \frac{\partial}{\partial y_ 0}
\end{eqnarray*}

By direct computation, but also form general considerations, these
vector fields generate the Lie algebra of $GL(3,\C)$. This algebra contains
Hamiltonian vector fields and vector fields which describe dissipation or
decoherence. These results are related to those of Rajeev (\cite{Rajeev:2007}).

\subsection{The analysis of entanglement}

Entanglement is a property of composite physical systems which plays a very
important role in many different phenomena, but in particular, it has become a
crucial issue of quantum computation and quantum information theory. Despite
the growing interest in recent years, it was already discussed by Schr\"odinger
and the ``founding fathers'' of quantum theory in the early years of quantum
theory (\cite{Schr:1935,Schr:1936}). 

Roughly speaking, entanglement is the situation complementary to separability

\begin{definition}
Let $|\psi\rangle$ be state of a Hilbert space $\mathcal{H}=\mathcal{H}_1\otimes \mathcal{H}_2$
of a bipartite system. Then, $|\psi\rangle$ is said to be {\bf separable} if there
exists a pair of states $|\psi_1\rangle\in \mathcal{H}_1$ and
$|\psi_2\rangle\in \mathcal{H}_2$ 
satisfying $|\psi\rangle=|\psi_1\rangle\otimes |\psi_2\rangle$.
\end{definition}

Thus a state which is not separable is said to be entangled. But
entanglement exhibits many interesting properties, for instance the fact that
there is a gradation in the level of entanglement of the different states. Thus
we can measure the entanglement of a state by using physical quantities. These
different observables are called {\bf entanglement witnesses}.

Usual choices to describe entanglement are the
{\bf concurrence} of the state, the {\bf von Neumann entropy} of one of its
partial traces (i.e. the entropy of the density state $\rho_1=\Tr_2\rho_\psi$ or
of $\rho_2=\Tr_1\rho_\psi$, where $\rho_\psi=|\psi\rangle\langle\psi|$).

\begin{definition}
   The concurrence of a density matrix $\rho\in \mathcal{D}(\mathcal{H})$ is defined
   as 
$$
C(\rho)=\mathrm{max}(0, 2\lambda_{max}(\hat \rho)-\Tr (\hat \rho))
$$
where $\hat \rho$ corresponds to 
$$
\hat \rho=\sqrt{ (\sigma_2\otimes \sigma_2)\rho^*(\sigma_2\otimes \sigma_2)\rho}
$$
and $\lambda_{max}(\hat \rho)$ stands for its largest eigenvalue.

\end{definition}

\begin{definition}
The  von Neumann entropy of a density matrix $\rho\in \mathcal{D}(\mathcal{H})$ is
defined as 
\begin{equation}
  \label{eq:entropy}
  S(\rho)=\Tr \rho\log (\rho).
\end{equation}
When the density matrix corresponds to a pure state, though, the function above
vanishes. Thus we define the corresponding entropy as the value of the function
on the partial trace over one of the subsystems:
\begin{equation}
  \label{eq:entropy_pure}
  S(\rho_\psi)=\Tr \rho_1\log (\rho_1) \qquad \rho_1=\Tr_1 \rho_\psi
\end{equation}
\end{definition}

On the other hand, if we consider the case of mixed states, the situation is
not equally simple. In general, it is necessary to consider more than one
entanglement witnesses in order to completely characterize the state of the
system. One interesting question arises thus: how can we describe the
independence of the different quantities we use? 

In the framework of classical mechanics this question is simple to
answer. Given two physical quantities, which are represented by two functions
$f_1, f_2$ on phase-space, they are said to be independent at a point $p$ if their
exterior differentials satisfy
$$
(df_1\land df_2)(p)\neq 0
$$

The usual approach to Quantum Mechanics, in terms of Hilbert spaces or
$\C^*$--algebras does not allow a similar treatment of the analogous quantum
problem. We lack of a well developed non-commutative differential calculus which would
allow to define a
``non-commutative'' exterior differential and therefore to extend  the
previous definition to the quantum setting
\cite{MarViZam:2006,Segal1,Segal2,BiMarStern:2000}

However the geometrical formalism we introduced in the previous sections allows us
to look at the problem from a different perspective. Treating the quantum state
space as a real differential manifold, we do have a differential calculus at
our disposal: the usual differential calculus of real manifolds.

    Consider the Hilbert space $\mathcal{H}$ and an operator $A$. We
know that we  can associate with $A$ the quadratic function
$$
A\to f_A(\psi)=\langle\psi|A|\psi\rangle\qquad \psi\in \mathcal{H}.
$$

In the geometric description of Quantum Mechanics we read from the set
of quadratic functions the algebraic structures available on the set
of operators:
\begin{itemize}
\item the associative product of operators is translated into the
  non-local product $\star$,
\item the Lie algebra defined by the commutator is translated into the
  Poisson algebra defined by the tensor $\Lambda$
\item the Jordan algebra given by the anticommutator is translated
  into the Jordan algebra defined by the tensor $G$
\end{itemize}

But the geometric description also includes a pointwise algebra
$(f_A.f_B)(\psi)=f_A(\psi)f_B(\psi)$, which is commutative, and whose
differential calculus is the standard one. This is the algebraic structure with
respect to which we define the differential algebra we are interested in:

    \begin{definition}
Two observables $A$ and $B$ are said to be {\bf functionally independent} if their
associated functions satisfy
$$
df_A\land df_B\neq0  \text{   on a dense submanifold of $\mathcal{H}$ }
$$      
    \end{definition}

This is a condition which is simple to test and exhibits a clear
advantage offered by the geometric approach to the description of
quantum systems, for there is no simple analogue of this structure in
the usual framework.

Let us see how this works in a particular example where we can test
the independence of von Neumann entropy and the concurrence of our two
qubit system:

\begin{example}

Now we will test this formalism with a particular example. Consider
for instance the family of density states defined by the matrices: 
$$
\rho_t=
\left (
\begin{array}{cccc}
 0 & 0 & 0 & 0\\
0& a& \frac 12 c e^{i\phi},0 \\
0 &\frac 12 c e^{-i\phi}& b & 0 \\
0& 0& 0& 1-a-b
\end{array}
\right ),
$$
as it is considered in \cite{Derkacz2005}. Such a matrix represents a density
state provided that  
$$
0\leq a+b \leq 1 \quad 0\leq c\leq 1\quad 4ab\geq c
$$
This is clearly a $4$--dimensional submanifold $\mathcal{S}$ of
$\D(\Hil)$. We can take an adapted basis for it, considering the
matrices 
$$
\left (
 \begin{array}{cccc}
 0 & 0 & 0 & 0\\
0& 1& 0 & 0 \\
0 &0 & 0  & 0 \\
0& 0& 0& -1
 \end{array}
\right ),
\left (
 \begin{array}{cccc}
 0 & 0 & 0 & 0\\
0& 0& 0 &0 \\
0 &0 & 1  & 0 \\
0& 0& 0& -1
 \end{array}
\right ),
\left (
 \begin{array}{cccc}
 0 & 0 & 0 & 0\\
0& 0& 1&0 \\
0 &1 & 0  & 0 \\
0& 0& 0&0
 \end{array}
\right ),
\left (
 \begin{array}{cccc}
 0 & 0 & 0 & 0\\
0& 0& i& 0 \\
0 &-i & 0  & 0 \\
0& 0& 0& 0
 \end{array}
\right )
$$

We can use the four real numbers $\{a,b,c,\phi\}$ as adapted coordinates on that submanifold.

Now we can evaluate the three functions above on these states. As we
already know the expression of the Jordan and the Poisson bracket we
can also obtain the corresponding Hamiltonian and gradient vector
fields. And besides, we can also study the independence of the
functions by evaluating the expressions of  
$$
 dS\land dC 
$$
where $d$ represents the exterior differential associated with the
differential manifold constituted by $\u^*(4)$. 

Let us thus proceed:

\begin{itemize}
 \item The value of the different functions is easy to obtain. We have
that von Neumann entropy reads,
\begin{eqnarray}
 2S(\rho_t)&=&-2 (-1+a+b) \log[1-a-b]+\nonumber \\ 
&&\left(a+b-\sqrt{(a-b)^2+c^2}\right)
\log \left[\frac{1}{2} \left(a+b-\sqrt{(a-b)^2+c^2}\right)\right]+
\nonumber \\
&&\left(a+b+\sqrt{(a-b)^2+c^2}\right) \nonumber
\log\left[\frac{1}{2} \left(a+b+\sqrt{(a-b)^2+c^2}\right)\right].
\end{eqnarray}

The value of the concurrence is very simple:
\begin{equation}
 C(\rho_t)=c
\end{equation} 

\item We can study now the Poisson brackets corresponding to them. It
  is simple to prove that these functions commute, i.e. 
\begin{equation}
 \{S,C\}=0
\end{equation} 
This implies that the local transformations generated by them  commute.
\item Finally, we can study the independence of the different
  functions. This is an important issue, in particular the
  independence of the von Neumann entropy and the concurrence, because
  it affects the description of entanglement of general quantum
  density states. We can prove the following: 
\begin{lemma}
 The concurrence and the von Neumann entropy of the family of states
 $\rho_t$ are not independent in all the space of density states. 
\end{lemma}

\begin{proof}
We are considering the submanifold of $\u(\Hil)$ corresponding to the family of
density states $\rho_t$.  

 Thus 
 the differential of the concurrence is trivial to obtain:
\begin{equation}
\label{wq}
 dC(\rho_t)=dc.
\end{equation} 
The computation of the differential of the von Neumann entropy is
quite more involved. It is evident from the expression above that the
functions $\mathcal{S}$ depends on the three variables. But as $C$
depends only on $c$, we have to consider only the $a$ and $b$
dependence in what regards the computation of $dS\land dC$. To this
aim we compute: 

\begin{eqnarray*}
\frac{\partial S}{\partial a}, \quad \frac{\partial S}{\partial b}
\end{eqnarray*}

Now, the condition for (\ref{wq}) to be equal to zero corresponds to 
$$
  \frac{\partial S}{\partial a} =0=\frac{\partial S}{\partial b} 
$$

And these conditions become
$$
 2 \text{Log}[1-a-b]+\text{Log}\left[a b-\frac{c^2}{4}\right]=0 \quad \text{Log}\left[4 a b-c^2\right]-2 \text{Log}\left[a+b+\sqrt{(a-b)^2+c^2}\right]=0
$$

These equations have a solution on 
$$
 \frac{1}{3}<a<\frac{1}{2};\qquad  b=a;\qquad c=\sqrt{-4+8 a-4 a^2+8 b-4 a b-4 b^2}
$$

If we represent the condition for $c$ as a function of $a$ and $b$ we
verify that it is well defined for all values of $a$ and $b$ (we
represent the function $c=\sqrt{-4+8 a-4 a^2+8 b-4 a b-4 b^2}$ and the
function $c=\sqrt{-a+16a-12a^2}$, which corresponds to the evaluation
on the submanifold $a=b$). 

\includegraphics{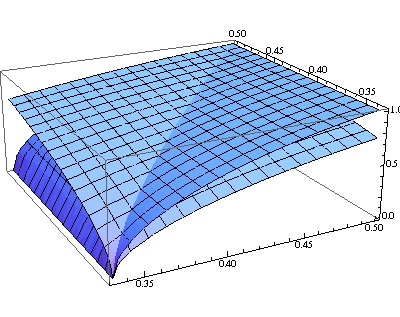}

Thus we conclude that there is a nonempty subset of $\u^*(4)$ where
the von Neumann entropy functions and the concurrence function
introduced above are not functionally independent. On any point outside this
submanifold the two functions are indeed functionally independent.  
\end{proof}

\end{itemize}
\end{example}

\section{Conclusions and outlook}

To summarize the exposition we would like to mention a few advantages and some
potential applications of the geometric approach to Quantum Mechanics that we
just presented to the field of Quantum Information Theory and Quantum Control. 

The main idea the paper aims to transmit is that Quantum Mechanics can be
efficiently presented as a geometric theory, considering that a quantum
dynamical system can be defined as
\begin{itemize}
\item a dynamical system on a K\"ahler manifold where the dynamical vector field
  is Hamiltonian with respect to the canonical symplectic structure and is a
  Killing vector for the Riemannian metric.
\item a dynamical system defined as a derivation for an associative algebra
  endowed with a pair   of operations which define a Lie-Jordan algebra on it. 
\end{itemize}

In both cases, the message is simple: the geometric framework provides an
approach to quantum mechanics which is very close to the usual geometric
framework of classical mechanics and many of the ``classical tools'' can
therefore be used in the quantum domain. The most
relevant is perhaps the differential structure encoded in the exterior
differential calculus. Such a structure has not a simple analogue in the usual
quantum description, but provides a powerful tool to study important notions as
the independence of operators, as we saw in the last section. Also the passage
to the projective space turns out to be a simple task for, up to some conformal
factors whose meaning is clear in geometrical terms, the treatment of the
projective theory is analogous to the one defined at the level of Hilbert spaces.

But the analogies of the quantum and the classical frameworks have also some
other deeper implications, particularly in the realm of Quantum Computation and
Quantum Information theory. We have seen how the cases of one and two qubits
and one qutrit can be perfectly described within the geometric approach in
fairly simple terms. The geometric implications of entanglement is under study
now, and we hope to be able to provide a detailed description in a short time.
But also quantum channels, their capacities, the study of decoherence,  
etc are concepts with a simple translation in the geometric domain. We are
presently addressing these issues and they will appear elsewhere.

Finally we would like to mention another advantage of the geometric description
of quantum mechanics: the study of quantum control problems. Geometric control
theory is a well stablished theory which has provided very powerful tools to
describe classical control systems. On the other hand, the usual description of
quantum control problems makes difficult to translate or adapt those classical
tools to the case of quantum systems. But once we have proved that quantum
dynamical systems can be described using the same mathematics which is used in the
classical case, the possibilities for application of classical geometric
control theory seem to be quite rich. And this is even true for systems and
phenomena purely quantum in origin, as the quantum Zeno effect, whose
applications to control problems are becoming an interesting topic
\cite{saverio,saverio2,FaMarPas:2007}.


\end{document}